\documentclass[twocolumn]{aastex701}

\usepackage{breqn}
\usepackage[caption=false]{subfig}
\usepackage{graphicx}
\usepackage[T1]{fontenc}
\usepackage{natbib}
\usepackage{amsmath}
\usepackage{float}

\begin{document}

\title{Spectral analysis of magnetized advective accretion flows around rotating black holes}

\author[0000-0002-4834-4211]{Mayank Pathak}
\affiliation{Joint Astronomy Programme, Department of Physics, Indian Institute of Science, Bengaluru 560012, India}
\email{mayankpathak@iisc.ac.in}

\author{Shubhrangshu Ghosh}
\affiliation{Center for Astrophysics, Gravitation and Cosmology (CAGC), Shri Ramasamy Memorial (SRM) University Sikkim, \\
5$^{\rm th}$ Mile Tadong, Gangtok, 737102, India} 
\affiliation{Department of Physics, Shri Ramasamy Memorial (SRM) University Sikkim, 5$^{\rm th}$ Mile Tadong, Gangtok, 737102, India} 
\email{shubhrangshughosh.r@srmus.edu.in} 

\author[0000-0002-3020-9513]{Banibrata Mukhopadhyay}
\affiliation{Joint Astronomy Programme, Department of Physics, Indian Institute of Science, Bengaluru 560012, India}
\affiliation{Department of Physics, Indian Institute of Science, Bengaluru 560012, India}
\email{bm@iisc.ac.in}

\begin{abstract}
The spectra of an accretion disk around black holes are the basic diagnostic tool to enlighten the underlying flows and then black holes.
Accretion flows around black holes, however, are controlled by parameters like the magnetic field, spin of the black hole, accretion rate and temperature of the flow. These quantities affect the (magneto)hydrodynamics of the flow thus consequently lead to variations in the spectrum. We first consider numerical steady state magnetohydrodynamic (MHD) solutions of magnetized accretion flows around black holes to study the dependence of the spectra on these disk properties. The spectrum exhibits strong dependence on the spin of the black hole, accretion rate, magnetic field and the electron temperature of the flow. Variations in these quantities influence the emission peaks and overall luminosity, which can be a tell-tale sign to extract physics of observed spectra. We further validate our results with general relativistic MHD (GRMHD) simulations using the standard and normal evolution (SANE) and magnetically arrested disk (MAD) vector potentials. We consider two black hole spins ($a=0.5$ and $a=0.9375$) to model the magnetic field configurations and study the resulting spectra by comparing MAD and SANE results. We find a large difference in the bolometric luminosities and the location of the emission peaks between SANE and MAD flows. Certain properties of the spectra, like, the ratio of synchrotron radiation to synchrotron self-Comptonization peaks in SANE and MAD, show drastically distinct features. The overall luminosity combined with such metrics can distinguish the magnetic field characteristics in astrophysical systems. 
\end{abstract}
\section{Introduction}
Spectra of astronomical objects have been a powerful tool in determining the inner working of these systems. X-ray binary accretion disks have famously been modeled with the ``thin-disk" model proposed by \cite{ss73} (SS hereafter). The emission is modeled by an optically thick, geometrically thin accretion disk around a black hole, with each radius of the disk radiating black body emission, leading to a multicolor black body spectrum. However, sources emitting non-thermal and power-law photons cannot be explained using the SS model. These types of sources were understood to be hotter than the SS disk, but to model such optically thin flows using approaches similar to SS lead to thermal instabilities \citep{sle}. To handle such instabilities in apparently radiatively inefficient systems, while modeling a hotter accretion flow than SS, advection of the accreting plasma was introduced. This stabilized the accretion system as the excess heat generated was quickly advected into the black hole. \cite{slim} proposed the slim disk model for optically thick advective accretion flow and showed the stability of the same at temperatures higher than the SS disk. \cite{tcaf} and \cite{adaf} explored the optically thin side of the parameter space for advective accretion flows and proposed the two component accretion flow (TCAF) and advection dominated accretion flow (ADAF) models, respectively, for low density (low accretion rate) flows around black holes. These models were able to achieve electron temperatures of $T_e\sim 10^9\text{K}$ and thus could explain the non-thermal and power-law emissions from accretion sources. However, the treatment of the hydrodynamics in these models was a bit simplistic, with no explicit solutions of magnetic field. In other words, any magnetic effects therein were assumed to be weak or of equipartition, without solving the Maxwell's equation along with.

The accretion flow is, however, dependent on several disk properties explicitly, like the magnetic field strength and geometry, spin of the black hole, accretion rate (determining the flow density), temperature etc. The susceptibility of the resulting spectra is also consequently dependent on these quantities, though these dependencies are ill understood, especially in accretion flows with strong magnetic fields.

In this work, we solve the magnetohydrodynamic (MHD) equations to construct a sub-Keplerian accretion flow around a rotating black hole. We explore the variations of spectra with the aforementioned disk properties to systematically understand the behavior/variation of each emission component from our sub-Keplerian accretion flow. 

We model the primary emission as a combination of synchrotron and bremsstrahlung radiations from the sub-Keplerian region, and soft photons from the geometrically thin disk (a multicolor blackbody spectrum). All three of these seed photon populations are then upscattered via inverse-Comptonization by the hot electrons in the sub-Keplerian plasma. As a first approximation, we restrict ourselves assuming a fully thermalized (Maxwellian) electron distribution. However, in reality, for hot advective accretion flows, processes like magnetic reconnection, turbulent dissipation, shock acceleration, etc., are all likely to accelerate a fraction of the ions and electrons producing a non-thermal population that follows a power-law energy distribution, eventually forming a hybrid thermal-nonthermal energy distribution (see, e.g., \citealt{mandal, yuan-2014}). 

To further validate the results obtained from our steady state accretion flow solutions, we use general relativistic magnetohydrodynamic (GRMHD) simulations using \texttt{BHAC} to compute the emission spectra for accretion flow (excluding the jet) with different magnetic field configurations. We consider the standard and normal evolution (SANE) and magnetically arrested disk (MAD) vector potentials to model the initial magnetic field configurations in our simulations and the evolved flow is then used to calculate the spectra. Such a study is useful for understanding the importance of magnetic field evolution in the context of emissions from accretion flows. Similar work was undertaken by \cite{xie} by modeling magnetic fields based on MAD and SANE simulations and then solving the steady state MHD equations. \cite{sdb-2022} also used GRMHD simulations to compare spectra from SANE and MAD simulations. They utilized particle in cell (PIC) simulations to post process the temperature of the accretion flow in their simulation domain. To further explore energy dissipation in MADs, \cite{sbd-2024} used thermally cooled GRMHD simulations to study geometrically thin MADs ($0.03<$ aspect ratio $<0.3$). These disks showed quasi-Keplerian angular speeds and thermally driven winds that can feed the coronae around black hole accretion sources, leading to non-thermal emissions.

In our calculations, we use the GRMHD disk data directly from \texttt{BHAC} and the electron temperature is then calculated in post-processing to evaluate the radiative components of the spectra, assuming the latter weakly depend on flow dynamics.

The paper is organized as follows. In section \ref{num-setup}, we describe the numerical set-up of the MHD equations being solved and calculation of the spectrum. In section \ref{results}, we describe the dependence of the spectra on the various parameters described in section \ref{num-setup} and discuss its implications. In section \ref{grmhd}, we describe our simulation set-up and the calculation of electron temperature. Consequently, we study the spectra for MAD and SANE simulations and compare them for two black hole spins in the same section. We conclude the paper in section \ref{conclusion} with a summary of our results and the future work to be done.

\section{Numerical set-up} \label{num-setup}
\subsection{Accretion disk solution in steady-state}\label{basic-eq} 
We describe the steady-state, sub-Eddington, advective, magnetized accretion flows around black holes using the pseudo-Newtonian framework. The matter-energy transport is thus governed by advection ($dv/dr$, where $v$ is the radial velocity and $r$ is the radial distance from the black hole) with all flow variables being time-independent ($\partial/\partial t=0$). Sub-Eddington accretion rates ($\dot{M}<0.01\dot{M}_{Edd}$) lead to the flow being weakly radiatively efficient, i.e., the heat radiation $Q^-$ is not negligible but not determining the dynamics.
We consider length in units of $r_g=GM/c^2$, time in $r_g/c=GM/c^3$, in our calculations, where $M$ is the mass of the black hole, $c$ is the speed of light and $G$ is the gravitation constant. Also the accretion rate is defined in units of Eddington rate $\dot{M}_{\rm Edd}$.
Following previous literature (e.g. \citealt{adaf,manmoto,mandal}), as a first approximation, we restrict our analysis to 1.5-D accretion system. This implies that our system is axisymmetric and we consider no motion in the vertical ($z$) direction. Nevertheless, the spectra computed from our model could effectively describe low/hard state in X-ray binaries and even quiescent sources, which are sub-Eddington and advective, while also reducing the complexity of the equations to be solved. 
We also do not capture any effects arising of non-axisymmetry and vertical structure due to strong magnetic fields, as have been observed in GRMHD simulations. Such analysis is left for future work.


The vertically averaged MHD equations considered to study the accretion flow are as follows \citep{bm-kc}:
\begin{equation}
    \dot{M}=4\pi\rho Hvr, \label{eqmdot}
\end{equation}
\begin{equation}
    v\frac{dv}{dr}+\frac{1}{\rho}\frac{dP}{dr}-\frac{\lambda^3}{r^3}+F=\frac{1}{4\pi\rho}\left(B_r\frac{dB_r}{dr}+s_1\frac{B_zB_r}{H}-\frac{B_\phi^2}{r}\right),
\end{equation}
\begin{dmath}
    v\frac{d\lambda}{dr}=\frac{1}{r\rho}\frac{d}{dr}(r^2W_{r\phi})+\frac{r}{4\pi\rho}\left(B_r\frac{dB_\phi}{dr}+s_2\frac{B_zB_\phi}{H}+\frac{B_rB_\phi}{r}\right), \label{ang-mom}
\end{dmath}
\begin{equation}
    \frac{P}{\rho H}=\frac{FH}{r}-\frac{1}{4\pi\rho}\left(B_r\frac{dB_z}{dr}+s_3\frac{B_z^2}{H}\right),
\end{equation}
\begin{dmath}
    \frac{v}{\Gamma_3-1}\left(\frac{dP}{dr}-\frac{\Gamma_1P}{\rho}\frac{d\rho}{dr}\right)=Q^+-Q^-
    =Q_{vis}^++Q_{mag}^+-Q_{vis}^--Q_{mag}^-=f_{vis}Q^+_{vis}+f_{m}Q^+_{mag},
\end{dmath}
where $f_{vis}$ and $f_m$ are the fractions of viscous and Ohmic heating going to the ions, and
\begin{equation}
    Q_{vis}^{+}=\frac{\alpha(P+\rho v^2)\lambda}{r^2}=\frac{Q^-_{vis}}{1-f_{vis}},
\end{equation}
\begin{equation}
    Q_{mag}^{+}=\frac{J^2}{\sigma}=\frac{\alpha c_s h}{4\pi}(\nabla\times B)^2=\frac{Q^-_{mag}}{1-f_{m}}.
\end{equation}
$\Gamma_1$, $\Gamma_3$ indicate the polytropic indices depending on the gas and radiation content in the flow \citep{chandra}.
Other quantities/variables have their usual meaning, see, e.g., \citealt{rajesh10, bm-kc}.

Here we assume the variation of $v$ and specific angular momentum ($\lambda$) in the vertical direction to be zero. Due to flux freezing, if an accretion system involves strong outflows/significant activity in the vertical direction, magnetic fields show significant non-trivial behavior in the $z$ direction. As we neglect vertical motion ($v_z=0$) in our calculations, appreciable change in magnetic fields in $z$ direction is not expected. Thus, magnetic field components are assumed to have small variations in the vertical direction, such that $\partial/\partial z\rightarrow s_i/z\sim s_i/H$ with $i\equiv 1,2,3$, where $H$ is the vertical scale height. We choose $s_1=s_2=-0.5$. As we choose $v_z=0$, the induction equation and Maxwell's equations dictate $s_3=0$, i.e., there is no vertical variation in $B_z$ \citep{bm-kc}. $\dot{M}$ is the conserved mass accretion rate, $\rho$ is the mass density, $P$ is the total pressure including the magnetic contribution ($P=P_{gas}+B^2/8\pi$, where $P_{gas}$ is the gas pressure), $W_{r\phi}=\alpha(P+\rho v^2)$ is the modified viscous shearing stress following the SS prescription \citep{ss73}. The origin of this $\alpha$-viscosity is attributed to small-scale magnetic turbulence driven by the magnetorotational instability (MRI) \citep{mri1,mri2}. The average effect of MRI is encapsulated by the $W_{r\phi}$ stress tensor, as defined above. $F$ is the gravitational force corresponding to the pseudo-Newtonian potential for Kerr black holes \citep{m02}, given by

\begin{equation}
    F=\frac{(r^2-2a\sqrt{r}+a^2)^2}{r^3(\sqrt{r}(r-2)+a)^2}.
\end{equation}

Upon solving the above coupled differential equations, we obtain,

\begin{equation}
    \frac{dv}{dr}=\frac{N}{D},
\end{equation}
where $N$ and $D$ are functions of the flow variables. Due to the transonic nature of the flow, $D$ becomes zero at some critical radius/point ($r=r_c$) in the accretion flow. For the accretion to continue without any discontinuities, $N$ must be zero at $r=r_c$ as well \citep{rajesh10}. The location $r=r_c$ is an important boundary and we provide all the conditions required for the solution of the flow equations at this critical point.
\begin{figure*}
\centering
\includegraphics[width=\textwidth]{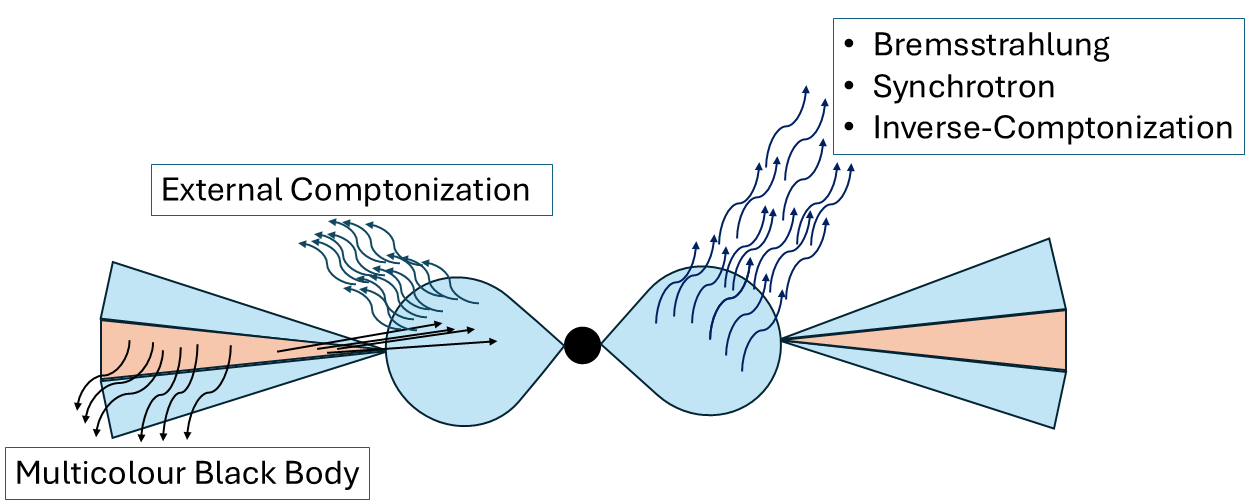}
     \caption{Schematic diagram of the model accretion system and radiation mechanisms operating in each part of the flow. The outer two-component region consists of the Keplerian (middle pink) and sub-Keplerian (enveloping sky-blue) disks and inner (quasi-spherical sky-blue) region is the pure sub-Keplerian disk. The non-thermal emission generated from the inner sub-Keplerian region and thermal emission from the Keplerian region are shown. The straight black arrows from the Keplerian (pink) region to the inner sub-Keplerian region are soft photons which get upscattered into the external Comptonization component. The radial extent of the sub-Keplerian region is subject to system parameters like the magnetic field, disk inner edge angular momentum etc. Similarly the size of the Keplerian region can be fixed by analyzing the observed multi-color black body spectra.}
     \label{model}
\end{figure*}

\subsubsection{Electron temperature calculation} \label{elec}

Accreting plasma is energized by both turbulent viscous dissipation and Ohmic heating. Typically, it has been assumed that the 
turbulent viscous heating is mainly associated with ions (as ions being more massive than electrons). However, it has been pointed out before that if magnetic field strength in the accretion flow is not 
substantially weak ($\beta_m=P_{gas}/P_{mag}\leq5$, where $\beta_m$ is the plasma-$\beta$ and $P_{mag}=B^2/8\pi$ is the magnetic pressure \citep{quat-1999}), electrons may receive direct heating (apart from Coulomb transfer of heat from ions) due to, like, Alfv\'enic turbulence, or pressure anisotropy in collisionless plasma, or turbulent magnetic reconnection of field lines in the accretion flow, mainly through the effect of Joules dissipation (see, e.g., \citealt{bk-97}; \citealt{quat-1999}; \citealt{sharma-2007}; \citealt{bk-1997}). Also in this scenario, it has been indicated that the electrons may acquire a share of viscous heating, even comparable to that received by ions, with the fraction of direct heating ranging in $\sim 0.1 - 0.5$ (see \citealt{xie-2012}). In fact, in their detailed modeling of 
Sgr A$^*$, \cite{yuan-2003} found that a viscous heating fraction of approximately $\sim 0.5$ is required to account for the observations. Here, in the present analysis, we consider that electrons directly gain energy from both turbulent viscous and Ohmic heating, with a maximum of viscous fraction $\sim 0.5$. The maximum fraction of turbulent Ohmic heating is assumed to be going to electrons is $\sim 0.9$, when it is being distributed among both species.  

We solve the electron energy equation to calculate the electron temperature in our accretion flow \citep{rajesh10}, given by

\begin{dmath}
    \frac{v}{\Gamma_3-1}\left(\frac{dP_e}{dr}-\frac{\Gamma_1P_e}{\rho}\frac{d\rho}{dr}\right)=Q_{vis}^{e+}+Q_{mag}^{e+}+Q^{ie}-Q^{e-}\label{sol-te},
\end{dmath}
where $Q_{vis}^{e+}=Q_{vis}^-$, $Q_{mag}^{e+}=Q_{mag}^-$ respectively describe the heating of the electrons via viscous and Ohmic heating, $P_e=\rho T_e$ and $T_e$ is the electron temperature \citep{rajesh10,bk-97}.
The electrons are cooled by bremsstrahlung, synchrotron radiations and their respective inverse-Comptonization (IC), given by
\begin{equation}
    q^{e-}=Q^{e-}c^{11}/(G^4M^3)=q^-_{br}+q^-_{syn}+q^-_{br,C}+q^-_{syn,C},
\end{equation}
where
\begin{multline}
    q_{br}^-=1.4\times10^{-27}n_in_e T_e^{1/2}(1+4.4\times10^{-10}T_e)\\
    \text{ erg cm}^{-3}\text{ s}^{-1},
\end{multline}
\begin{equation}
    q_{syn}^-=\frac{2\pi}{3c^2}kT_e\frac{\nu_c^3}{R}\text{ erg cm}^{-3}\text{ s}^{-1},
\end{equation}
\begin{dmath}
    q_{br,C}^-=3\eta_1q_{br}^-\left\{\left(\frac{1}{3}-\frac{x_c}{3\theta_e}\right)-\frac{1}{\eta_2+1}\left[\left(\frac{1}{3}\right)^{\eta_2+1}-\left(\frac{x_c}{3\theta_e}\right)^{\eta_2+1}\right]\right\}\text{ erg cm}^{-3}\text{ s}^{-1}, \label{brc}
\end{dmath}
\begin{equation}
    q_{syn,C}^-=q_{syn}^-\eta_1\left[1-\left(\frac{x_c}{3\theta_e}\right)^{\eta_2}\right]\text{ erg cm}^{-3}\text{ s}^{-1}. \label{syncc}
\end{equation}
Here $n_i$ and $n_e$ are the number densities of protons/ions and electrons respectively, $\theta_e=kT_e/m_ec^2$,
$x_c=h\nu_c/(m_ec^2)$ and $\nu_c$ is the synchrotron self-absorption cut-off frequency, $\eta_1=p(A-1)/(1-pA)$, where $p=1-\exp(-\tau_{es})$ and $A=1+4\theta_e+16\theta_e^2$. $\eta_2=-[1+\ln(p)/\ln(A)]$.

The electron heating due to Coulomb coupling with the proton is given by \citep{ULX20},
\begin{multline}
    q^{ie}=8\sqrt{2\pi}e^4n_in_e\left(\frac{T_e}{m_e}+\frac{T_p}{m_p}\right)^{-3/2} \ln(\Lambda) (T_p-T_e) \\
    \text{ erg cm}^{-3}\text{ s}^{-1},
\end{multline}
where, $T_p$ is the ion temperature, the Coulomb logarithm, $\ln(\Lambda)\approx20$ and $q^{ie}=Q^{ie}c^{11}/(G^4M^3)$.

\subsection{Evaluation of spectrum} \label{spec-cal}
Here, we consider the unscattered flux from the accretion flow as given by \citep{manmoto},
\begin{equation}
    F_\nu=\frac{2\pi}{\sqrt{3}}B_\nu[1-\exp(-2\sqrt{3}\tau_\nu)], \label{flux}
\end{equation}
where $\tau_\nu=\sqrt{\pi}\kappa_\nu H/2$ is the optical depth and $\kappa_\nu=\chi_\nu/(4\pi B_\nu)$ is the absorption coefficient. $\chi_\nu=\chi_\nu^{brem}+\chi_\nu^{synch}$ is the total emissivity due to bremsstrahlung ($\chi_\nu^{brem}$) and synchrotron ($\chi_\nu^{synch}$) emissions, and $B_\nu$ is the Planck function. 

The IC upscattering defined in Eqs. (\ref{brc}) and (\ref{syncc}) are averaged over all frequency ranges. This does not capture the upscattering (increase in energy/frequency of photons) in the spectra. To capture this effect and its variations with system parameters, we apply the formulas of IC scattering probability and scattering rate, as suggested by \cite{coppi}. 

We adopt a two-component accretion flow model, depicted in Fig.~\ref{model}. A nonmagnetic version of similar accretion model, considering shock in the flow, was proposed by \cite{tcaf}. The shock discussed by \cite{tcaf} is of the Rankine-Hugoniot type, formed due to the centrifugal barrier in the in-spiraling flow. The post-shock region becomes compressed, hot and geometrically thicker than the pre-shock region leading to significant deviation from the SS prescription. The detailed spectra of such accretion flows were studied earlier by \cite{mandal}. Nevertheless, magnetic fields can lead to geometrically thick and hot accretion flows without the need of shock \citep{bm-kc,rn2022,fluxerr}. Although shocks can exist in magnetized accretion flows as well, we do not look into the issue of shock and its fate in the presence of strong magnetic fields considered here.
The sub-Keplerian disk close to the black hole is geometrically thicker and hotter than the Keplerian disk. The idea is, far away from the black hole the disk is Keplerian 
 as modeled by \cite{ss73} and is enveloped by hot sub-Keplerian components. As the matter advances, the thinner Keplerian disk disappears and the flow becomes purely thicker sub-Keplerian. The size of the sub-Keplerian region depends on system parameters, like, disk specific angular momentum particularly around the critical point, magnetic field in the accretion flow etc. They are also indirectly controlled by the spin of black hole. The size of the Keplerian region is also model-dependent and can be fixed by fitting the observed multi-color black body spectra with the appropriate disk size. We consider the outer edge of our Keplerian region to be at $r=200r_g$ for our parametric study. Soft photons from the Keplerian region travel through the hotter sub-Keplerian region and get scattered by IC, leading to an additional component in the spectrum. This is the external Comptonization (EC) contribution in addition to the IC peaks for synchrotron and bremsstrahlung cooling. The soft photons from the Keplerian disk are injected at the transition edge between the Keplerian and sub-Keplerian disks. The number density of soft photons ($n_s(r)$) is assumed to be distributed in the thick disk as $n_s(r)=n_{ss}r_t/(2r_t-r)^3$, where $n_{ss}$ is the number density of soft photons injected at the transition radius, $r_t$. This distribution is motivated by the inverse cubic radial dependence of photons in thin disks \citep{liv-rev}. EC scattering from the enveloping region over the Keplerian disk (outer sky-blue region in Fig.~\ref{model}) is not considered.
 As most of the radiation emitted in our calculations originates close to the black hole, we account for the gravitational red-shift by multiplying the emitted luminosity by $(1-r_h/r)^{1/2}$, where $r_h$ is the radius of the event horizon of the black hole. The effect of Doppler shift on the emitted luminosity due to gas motion has also been included by the term $1/[1-(v/c)^2]$, where $v$ is the magnitude of the flow velocity \citep{manmoto}.

\begin{figure*}
\centering
\subfloat[]{
\includegraphics[width=0.48\textwidth]{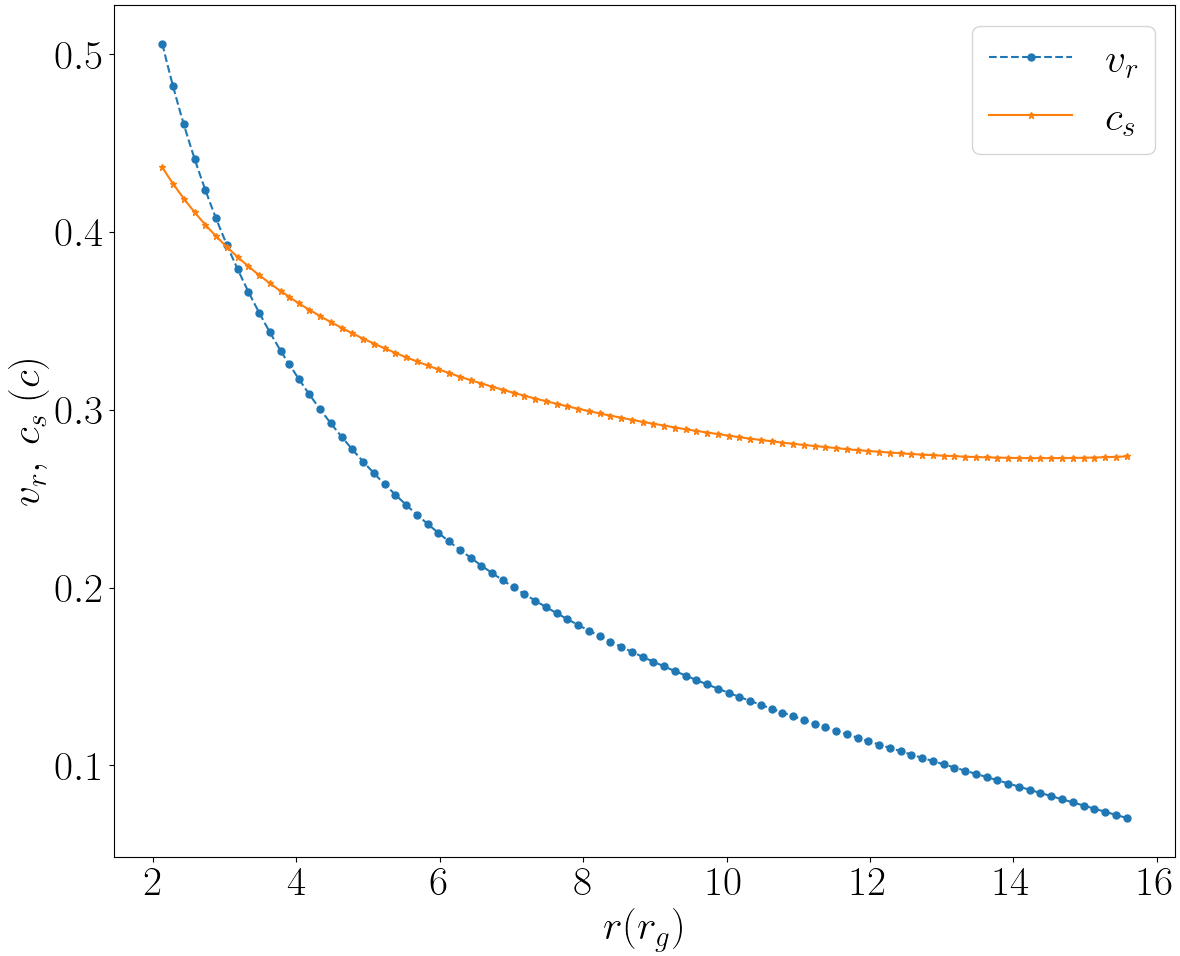}\label{vel}
}
\subfloat[]{
\includegraphics[width=0.49\textwidth]{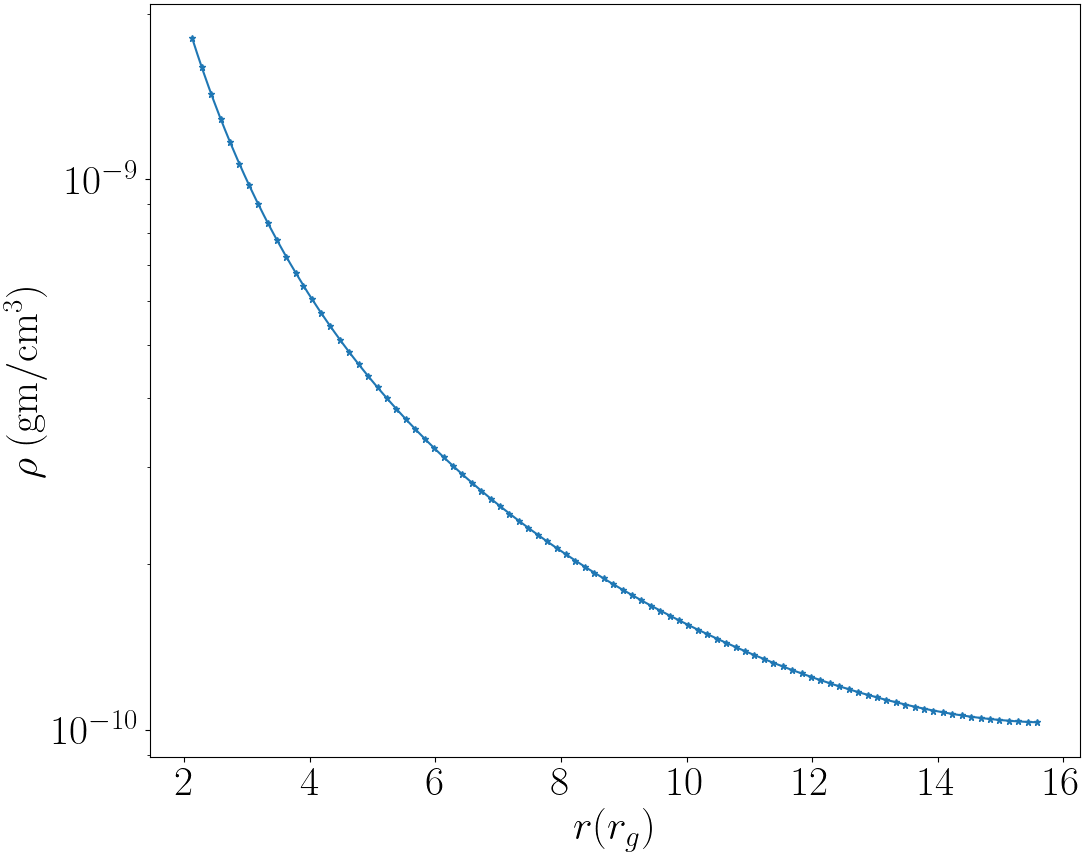}\label{rho}
}

\subfloat[]{
\includegraphics[width=0.49\textwidth]{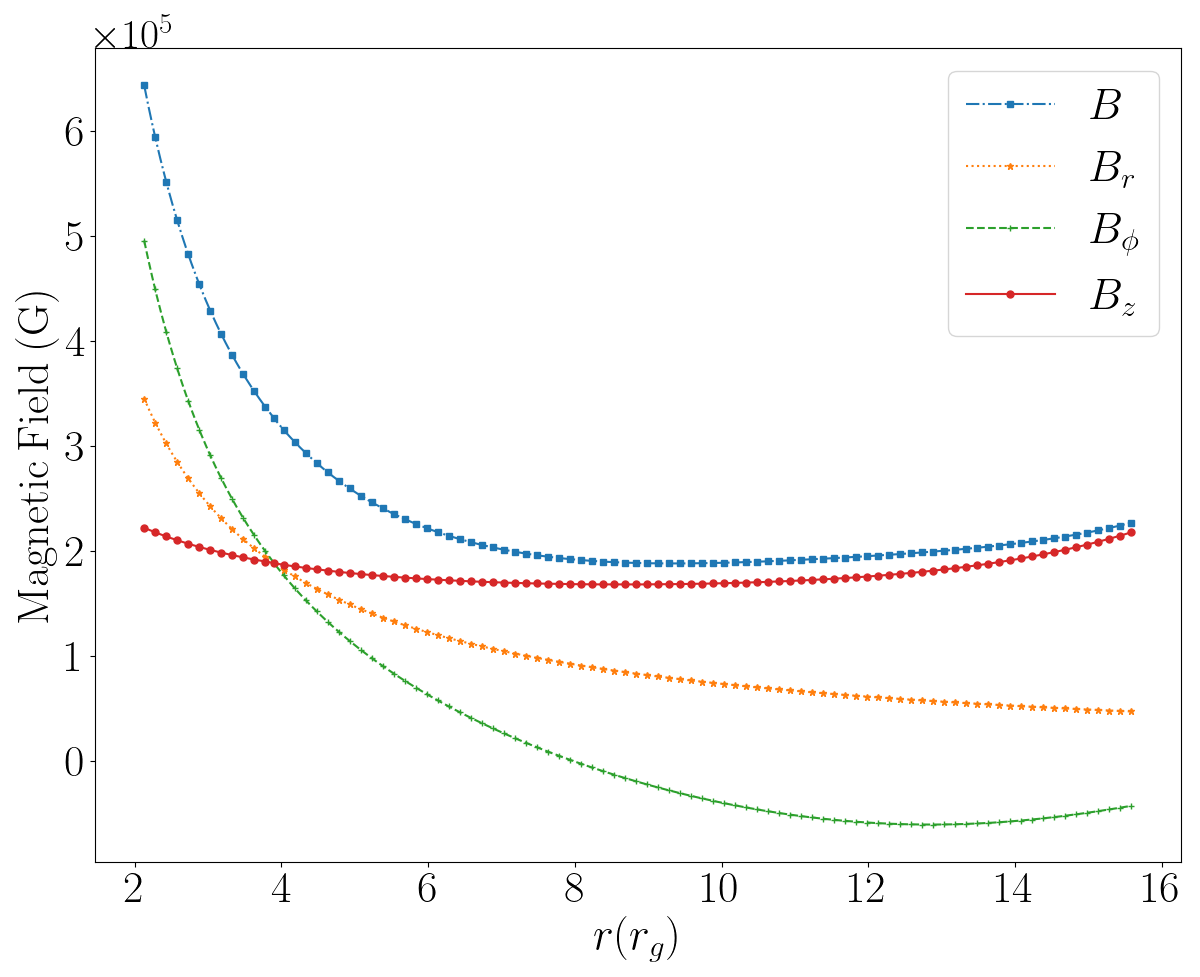}\label{mag}
}
\subfloat[]{
\includegraphics[width=0.475\textwidth]{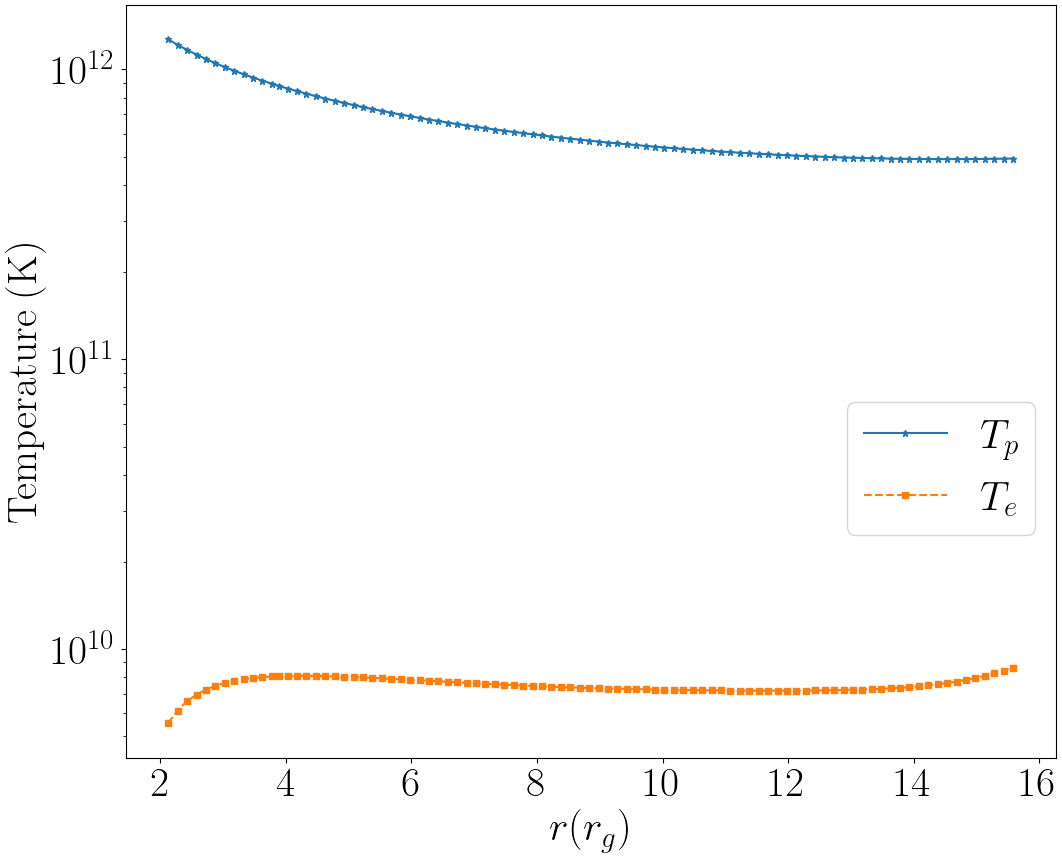}\label{temp}
}
     \caption{Flow profiles of (a) radial velocity and sound speed, (b) density, (c) magnetic field strength magnitude, and (d) ion and electron temperatures, for an accretion flow with $\dot{M}=10^{-3}$, $\alpha=0.01$, $M=10M_\odot$, $a=0.998$, $x_f=0.5$ and $x_{fm}=0.9$. 
     }
     \label{pro}
 \end{figure*}

\section{Results and Discussion} \label{results}

As mentioned in Sec. \ref{basic-eq}, our accretion system is described by several parameters namely, $\dot{M}$, $a$, $\alpha$, $x_f=1-f_{vis}$, $x_{fm}=1-f_m$, and magnetic field magnitude (B).

Flow profiles of relevant quantities for spectra calculations are given in Fig.~\ref{pro}. The radial velocity and sound speed decrease with radial coordinate from the black hole with a critical point around $r=3r_g$, showcasing the transonic nature of  accretion flow, with an X-type (saddle-type) critical point. As we are targeting sub-Keplerian flows, our flow density is $\sim10^{-9}-10^{-10}\mathrm{gm/cm^3}$. Fig. \ref{mag} shows the total magnetic field profile ($B$) with all the components. In this case our magnetic field shows non-monotonic characteristics, due to an increase in $B_z$ with increasing $r$. It is due to that away from the black hole, the flow remains magnetized but in the upper sub-Keplerian part, while the middle Keplerian part, by definition, is non- or weakly magnetic. However, the present work is confined to the inner pure sub-Keplerian flow, hence the evolution of magnetic field beyond the pure sub-Keplerian region has not been explored.

We also show $T_e$ and $T_p$ profiles in Fig.~\ref{temp}. $T_e$ is more than $100$ times less than $T_p$, which is expected in two-temperature accretion flows. 

\begin{figure*}
\centering
\subfloat[$M=10M_\odot$]{
\includegraphics[width=0.49\textwidth]{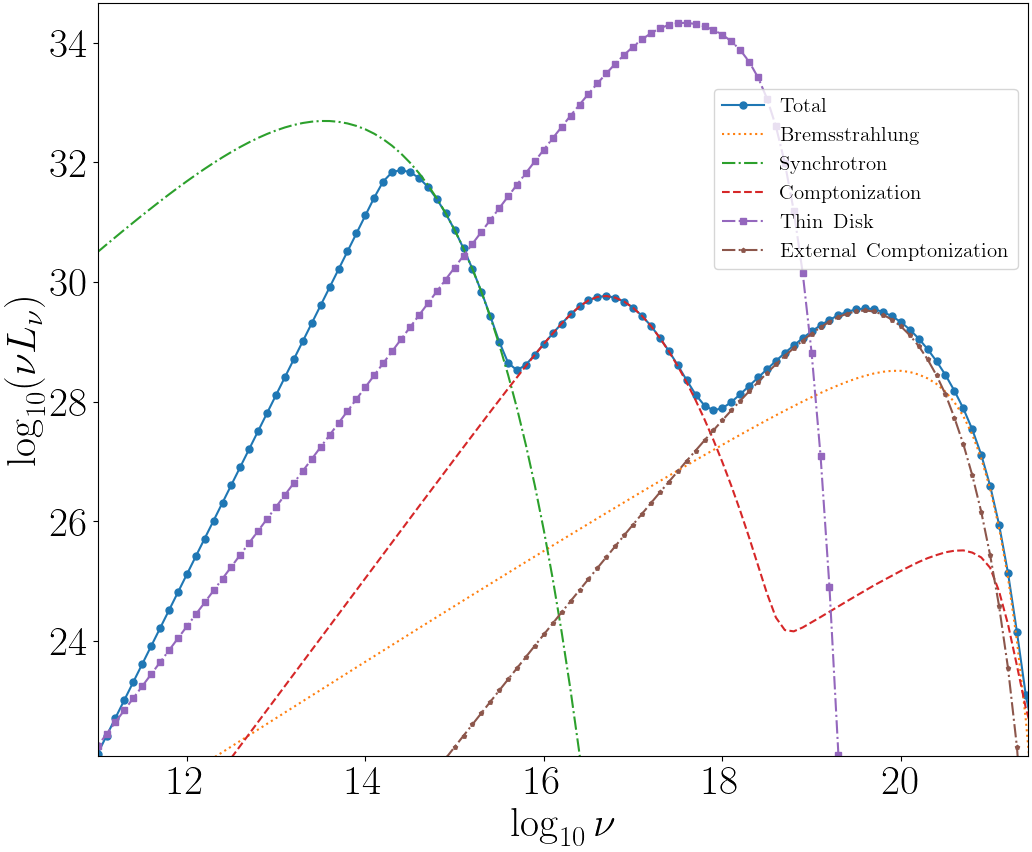}\label{sbh}}
\subfloat[$M=10^8M_\odot$]{
\includegraphics[width=0.49\textwidth]{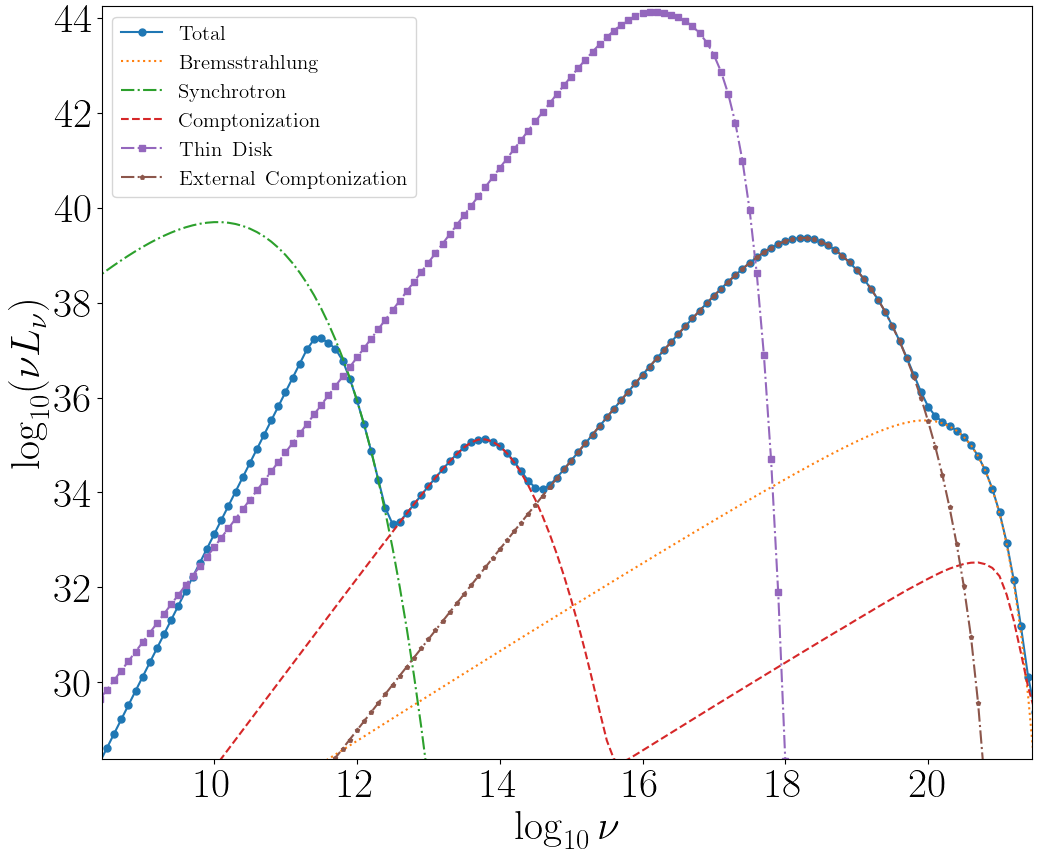}\label{smbh}
}
     \caption{Spectra of (a) stellar mass ($M=10M_\odot$), and (b) supermassive ($M=10^8M_\odot$) black holes with $\dot{M}=10^{-3}$, $\alpha=0.01$, $a=0.998$, $xf=0.5$ and $xfm=0.1$. Note the difference in the peak luminosities of the stellar mass and the supermassive cases.}
     \label{spec1}
 \end{figure*}

\begin{figure}
\centering
\includegraphics[width=0.47\textwidth]{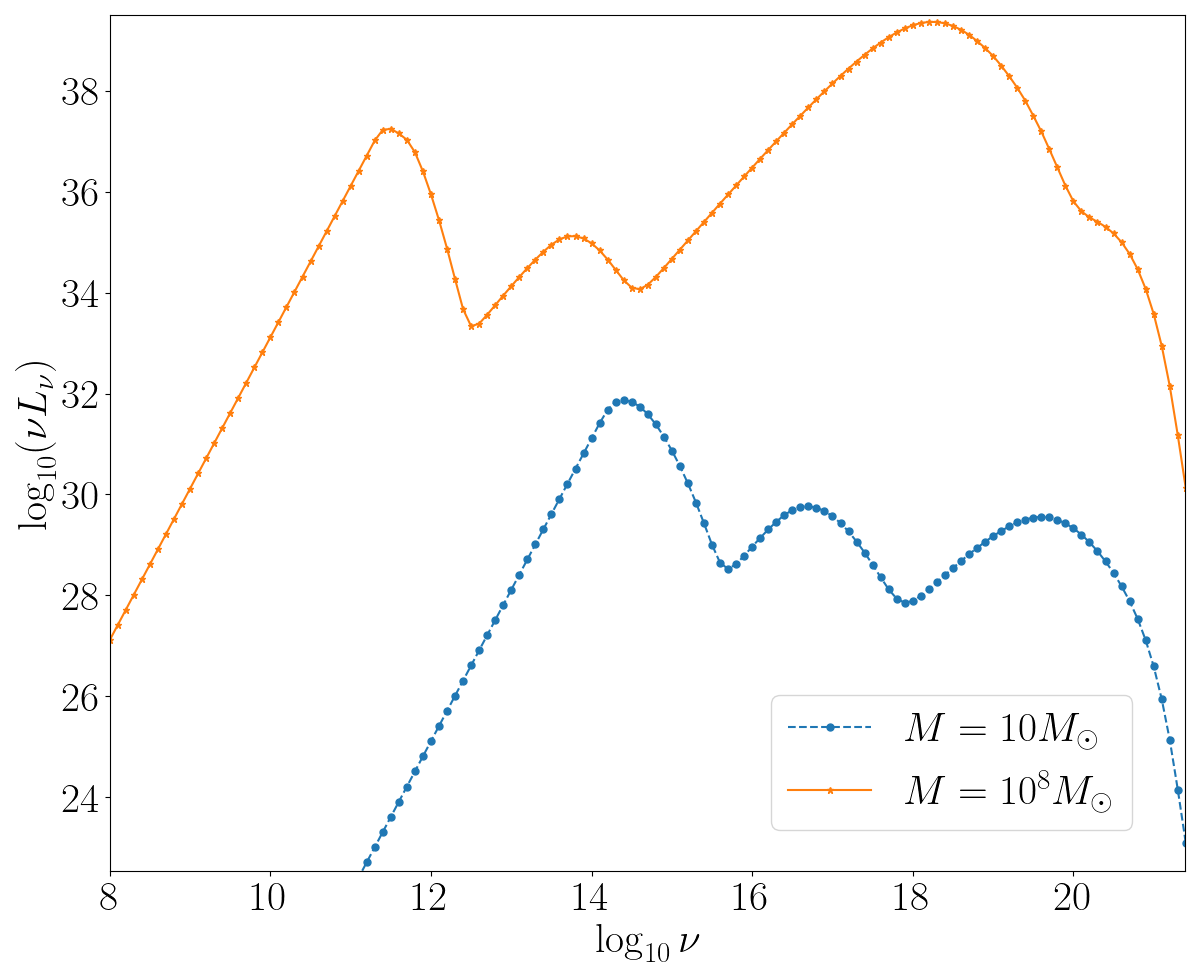}
     \caption{Comparison of stellar mass ($M=10_\odot$) and supermassive ($10^8M_\odot$) black hole spectra for $a=0.998$.}
     \label{spec-comp}
\end{figure}

\begin{table*}
\centering
\setlength{\tabcolsep}{8pt}
\begin{tabular}{c c c c c c c c c c c c c}
\hline
$a$ & $\dot{M}(\dot{M}_{Edd})$ & $r_c (r_g)$ & $\lambda_c (r_gc)$ & $T_{ec}(\mu_em_pc^2/k_b)$ & $\alpha$ & $x_f$ & $x_{fm}$ & disk size ($r_g$) & $m_i$ & $\beta_{mc}$\\
\hline
0.0 & $10^{-3}$ & 5.8 & 3.2 & $5.1\times10^{-4}$ & 0.01 & 0.5  & 0.9 & 13.89 & 8 & 88.62\\
0.5 & $10^{-3}$ & 4.8 & 2.6 & $5.9\times10^{-4}$ & 0.01 & 0.5  & 0.9 & 13.63 & 5 & 34.62\\
0.998 & $10^{-3}$ & 3.9 & 1.6 & $6.5\times10^{-4}$ & 0.01 & 0.5 & 0.9 & 13.46 & 3 & 12.46\\
0.998 & $10^{-3}$ & 3.9 & 1.6 & $6.5\times10^{-4}$ & 0.01 & 0.5 & 0.5 & 13.46 & 3 & 12.46\\
0.998 & $10^{-3}$ & 3.9 & 1.6 & $6.5\times10^{-4}$ & 0.01 & 0.5 & 0.1 & 13.46 & 3 & 12.46\\
0.998 & $10^{-3}$ & 3.9 & 1.6 & $1.8\times10^{-3}$ & 0.05 & 0.5 & 0.9 & 13.1 & 3 & 12.46\\
0.998 & $10^{-3}$ & 3.9 & 1.6 & $2.6\times10^{-3}$ & 0.1 & 0.5 & 0.9 & 12.61 & 3 & 12.46\\
0.998 & $10^{-3}$ & 3.9 & 1.6 & $6.5\times10^{-4}$ & 0.01 & 0.0 & 0.9 & 13.46 & 3 & 12.46\\
0.998 & $10^{-3}$ & 3.9 & 1.6 & $9.5\times10^{-4}$ & 0.01 & 0.5 & 0.9 & 13.46 & 3 & 12.46\\
0.998 & $10^{-3}$ & 3.9 & 1.6 & $1.6\times10^{-3}$ & 0.01 & 0.5 & 0.9 & 13.46 & 3 & 12.46\\
0.998 & $10^{-3}$ & 3.9 & 1.6 & $6.5\times10^{-4}$ & 0.01 & 0.5 & 0.9 & 24.73 & 30 & $1.25\times10^3$\\
0.998 & $10^{-3}$ & 3.9 & 1.6 & $6.5\times10^{-4}$ & 0.01 & 0.5 & 0.9 & 47.95 & 300 & $1.25\times10^5$\\
0.998 & $10^{-2}$ & 3.9 & 1.6 & $5.5\times10^{-4}$ & 0.01 & 0.5 & 0.9 & 13.46 & 3 & 12.46\\
0.998 & $5\times10^{-3}$ & 3.9 & 1.6 & $6.5\times10^{-4}$ & 0.01 & 0.5 & 0.9 & 13.46 & 3 & 12.46\\
0.998 & $10^{-4}$ & 3.9 & 1.6 & $6.5\times10^{-4}$ & 0.01 & 0.5 & 0.9 & 13.46 & 3 & 12.46\\
\hline
\end{tabular}
\caption{Disk parameters for the various cases considered under our study. Here temperature is defined in units of $\mu_em_pc^2/k_b\approx1.24\times10^{13}\mathrm{K}$, $\mu_e$ is the effective molecular weight of electrons, $m_p$ is the mass of protons and $k_b$ is the Boltzmann's constant. $m_i=m_r=m_\phi=m_z$. $\beta_{mc}$ is the plasma-$\beta$ at the critical point.}
\label{params}
\end{table*}

\subsection{Components of spectra for stellar mass and supermassive black holes}

Each annular region of the accretion disk exhibits emission due to the various radiative mechanisms. We calculate the emitted flux from each annular region and the total emission is calculated by summing up this emission over the entire disk.

To analyze the effect of various radiative processes leading to the shape of the spectrum from our accretion flow, we unfold various components of the spectra for stellar mass black hole (SBH) and supermassive black hole (SMBH) cases.

The corresponding spectra are shown in Fig. \ref{spec1}. We show the synchrotron radiation, bremsstrahlung, and their respective IC scattering spectral features. We also show the multicolor blackbody spectrum due to emission from the thin disk component which exists outside our sub-Keplerian disk. The blue-solid (with filled circle) line traces the total emission from the sub-Keplerian disk region (including the EC contribution), due to these processes. 

At lower frequencies, the spectra trace the blackbody branch (even corresponding to the hotter sub-Keplerian flow) for all cases. This is due to higher optical depth at these frequencies, leading to $F_\nu$ given in Eq.~(\ref{flux}) to reduce to the Planck function (modulo some constants). As the frequency increases, the optical depth decreases and the spectrum shifts to the optically thin branch, described by the emissivities of  non-thermal radiative processes. This point of transition marks the synchrotron peak in our spectra at around $\nu\sim10^{14}\mathrm{Hz}$ for SBH and around $\nu\sim10^{11}\mathrm{Hz}$ for the SMBH cases. This is because density in accretion flows scales inversely with the mass of the central black hole (from Eq.~\ref{eqmdot}, $\rho\sim\dot{M}/Hrv\sim\dot{M}/r_g^2c$; as $r_g$, $H$ (in terms of $r_g$) and $\dot{M}$ (in terms of $\dot{M}_{Edd}=1.38\times10^{17}M/M_\odot\hspace{0.05in}\mathrm{gm/s}$) are proportional to $M$, $\rho\sim1/M$). Thus, the SBH case has about $7$  orders of magnitude higher density and about $3$  orders of magnitude higher magnetic field than the SMBH case, leading to lower energy synchrotron photons for the latter. The synchrotron self-Comptonization (SSC) peak also shifts to lower energy due to less energetic scattering of lower energy photons for SMBH compared to SBH. The overall flux of the system, however, increases in the SMBH case as the size of the emission region scales as $M^2$, leading to much higher bolometric luminosity. These features are clearly evident in Fig.~\ref{spec-comp}.

The red-dashed line in Fig.~\ref{spec1} represent the IC scattering, showcasing two humps representing IC scattering of synchrotron radiation and bremsstrahlung. The peaks occur at higher frequencies than their respective seed photons due to upscattering of these photons by the hot electrons in the sub-Keplerian disk. A shift in three orders is observed for the IC for synchrotron radiation, but only an order or less for bremsstrahlung. The synchrotron IC peak forms a prominent emission peak in the X-ray regime while the bremsstrahlung IC occurs at $\gamma-$ray energies, albeit with low flux and does not affect the overall shape of the spectrum.

The EC component is upscattered by about two orders of magnitude from the thin disk component. For the SBH case, this component entirely covers the bremsstrahlung branch, while for SMBH it lies midway between the thin disk (``blue bump") contribution and bremsstrahlung branch. The luminosity due to this component is highest among all the non-thermal components, thus showcases the importance of EC contribution for black hole accretion spectra, especially for SMBHs. 

The bremsstrahlung peak occurs at the thermal frequency given by $\nu=k_bT_e/h$ ($\sim10^{20}$ Hz for our case) \citep{rn-nature}. Its position on the spectrum is thus governed solely by $T_e$ of the accretion flow. As $T_e$ does not vary much with the mass of the central black hole, the bremsstrahlung peak does not show any noticeable shift in its frequency for the SBH and SMBH cases.

\subsection{Parametric spectral variation} \label{para}
Our accretion depends on several parameters as described in \S\S~\ref{basic-eq}, \ref{results}. We show the variation of critical point electron temperature and size of the sub-Keplerian flow with  the change of these parameters in Table~\ref{params}. The magnetic field magnitude of each component is described by the ratio of sound speed to Alfv\'en speed at the critical point ($m_i=c_s/(B_i/\sqrt{4\pi\rho})$; $m_r, m_\phi$ and $m_z$ correspond to the $r$, $\theta$ and $\phi$ components of the magnetic field respectively). We consider an isotropic magnetic field at the critical point, i.e., $m_r=m_\phi=m_z$. The plasma-$\beta$ at the critical point ($\beta_{mc}$) is also shown in the last column. These values are chosen so that we get an X-type (saddle-type) sonic point for all the cases considered. For the $a=0.998$ case, the supplied $\beta_{mc}=12.46$, which is also observed in GRMHD simulations \citep{rn2022,chan15,mp-bm}.

\begin{figure}
\centering
\includegraphics[width=0.49\textwidth]{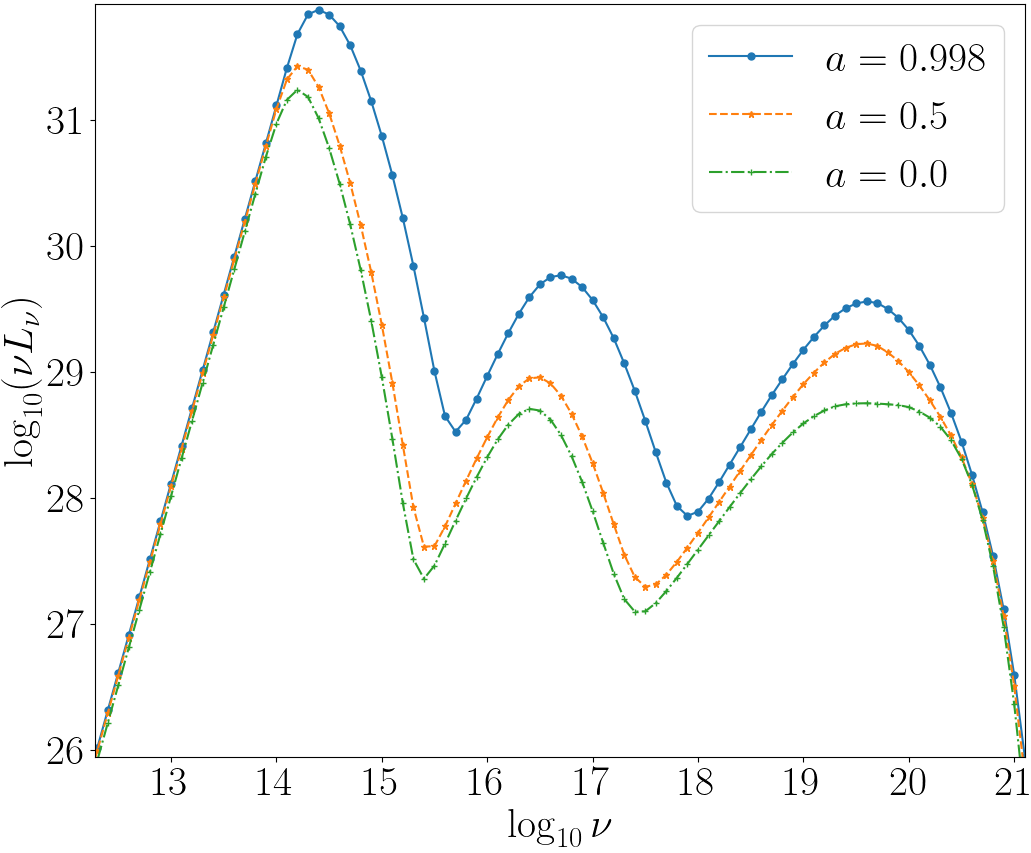}
     \caption{Variation of overall spectra with black hole spin $a$ for $M=10M_\odot$. The other parameters are same as in Fig.~\ref{spec1}. The first, second and third peaks are from synchrotron radiation, SSC and EC respectively.}
     \label{spec-a}
\end{figure}

 \begin{figure*}
\centering
\subfloat[]{
\includegraphics[width=0.47\textwidth]{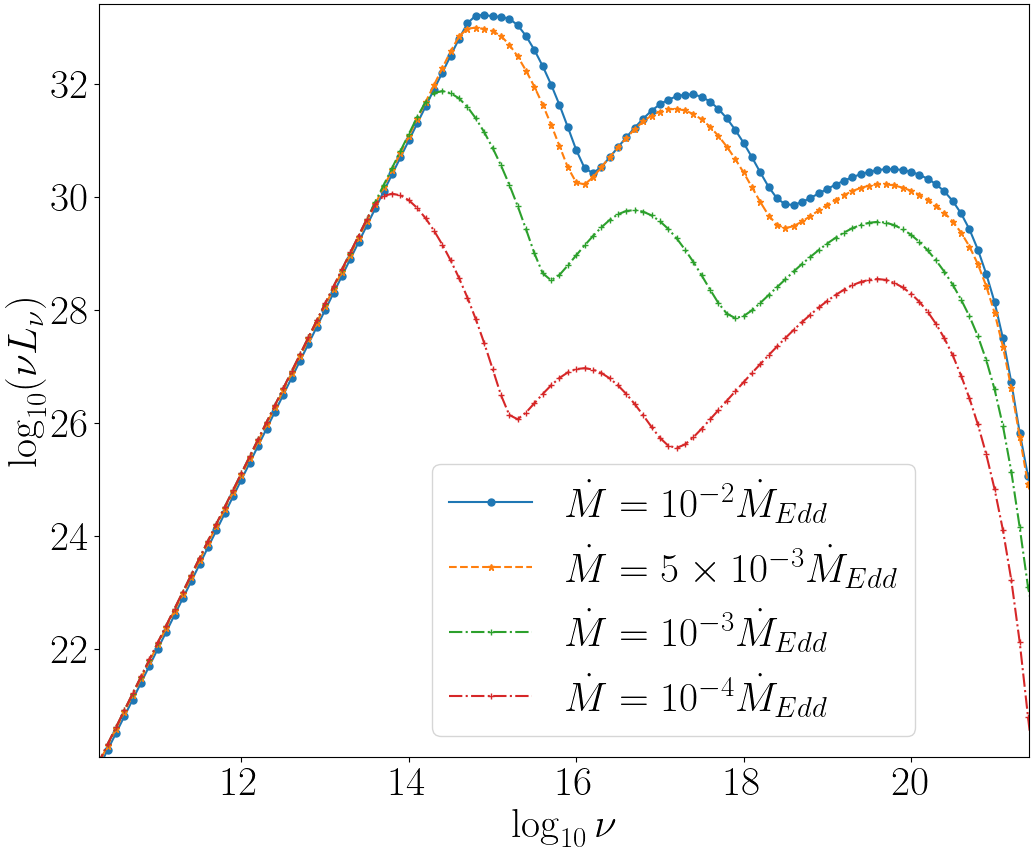}\label{s-mdot}}
\subfloat[]{
\includegraphics[width=0.49\textwidth]{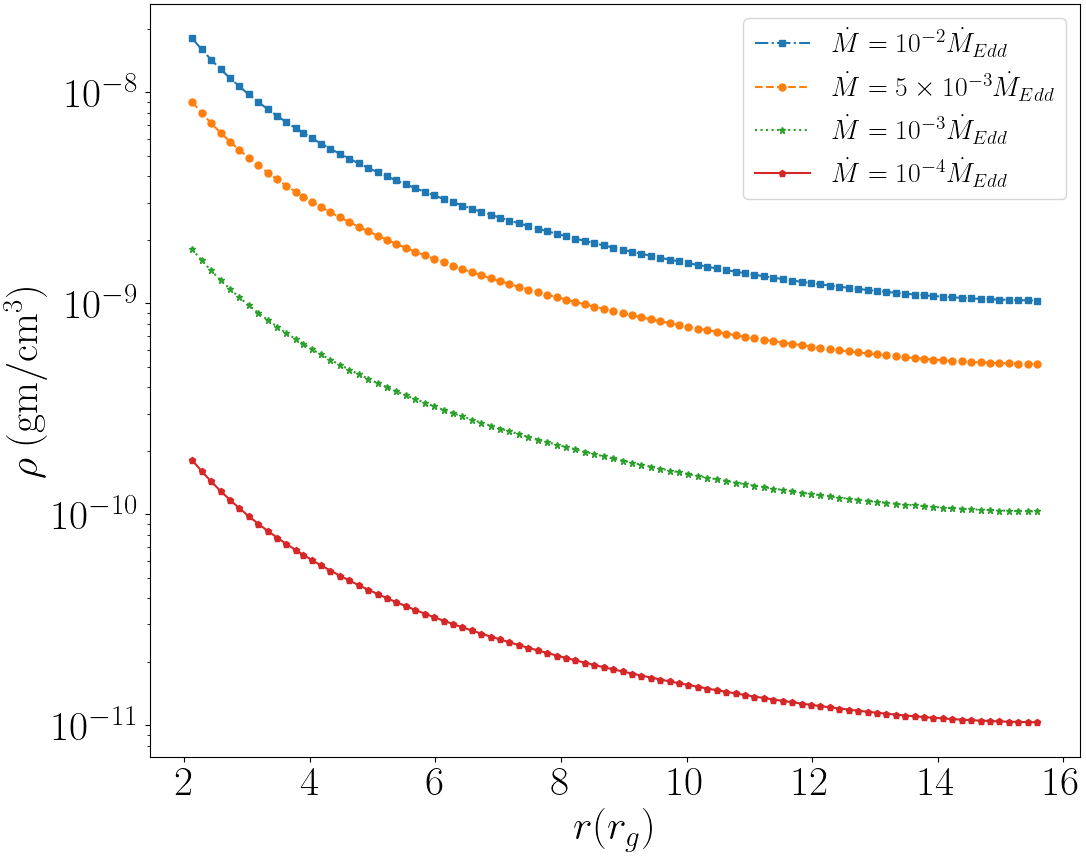}\label{a9-den}
}
     \caption{Variation of overall spectra and density profiles with accretion rate $\dot{M}$ for $M=10M_\odot$. The other parameters are same as in Fig.~\ref{spec1}. The first, second and third peaks in the spectra are from synchrotron radiation, SSC and EC respectively.}
     \label{spec-mdot}
 \end{figure*}

\begin{figure*}
\centering
\subfloat[]{
\includegraphics[width=0.48\textwidth]{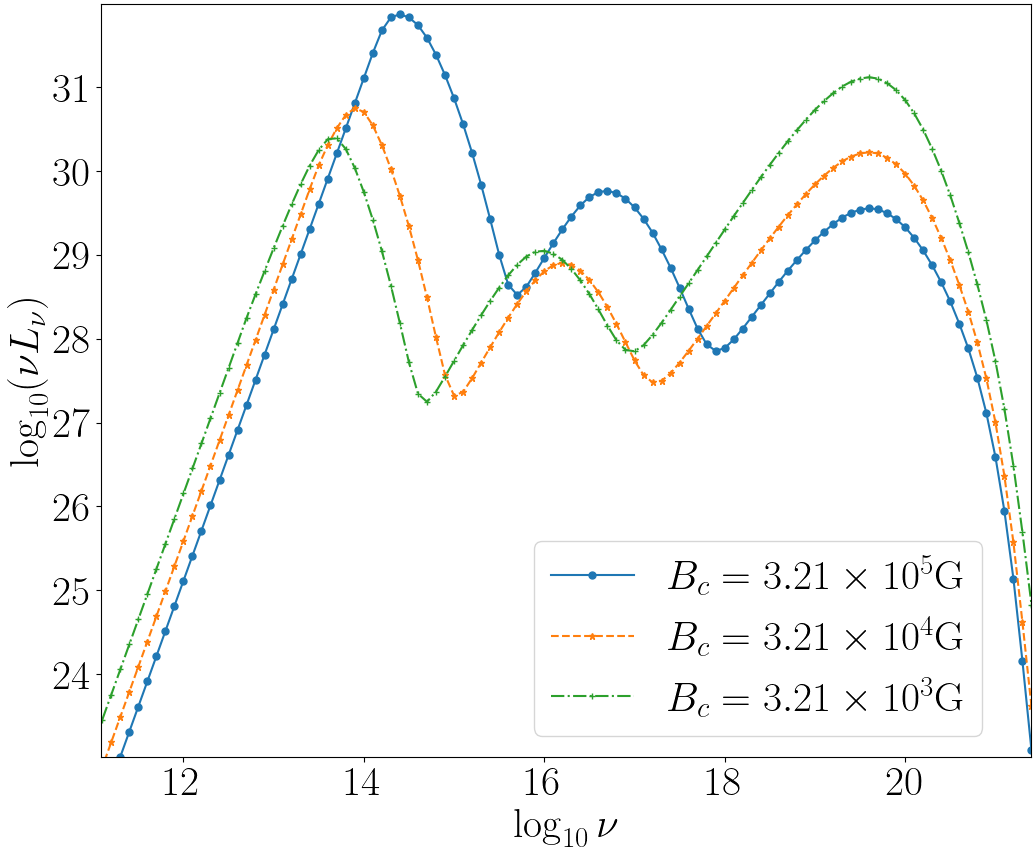}\label{s-b}}
\subfloat[]{
\includegraphics[width=0.485\textwidth]{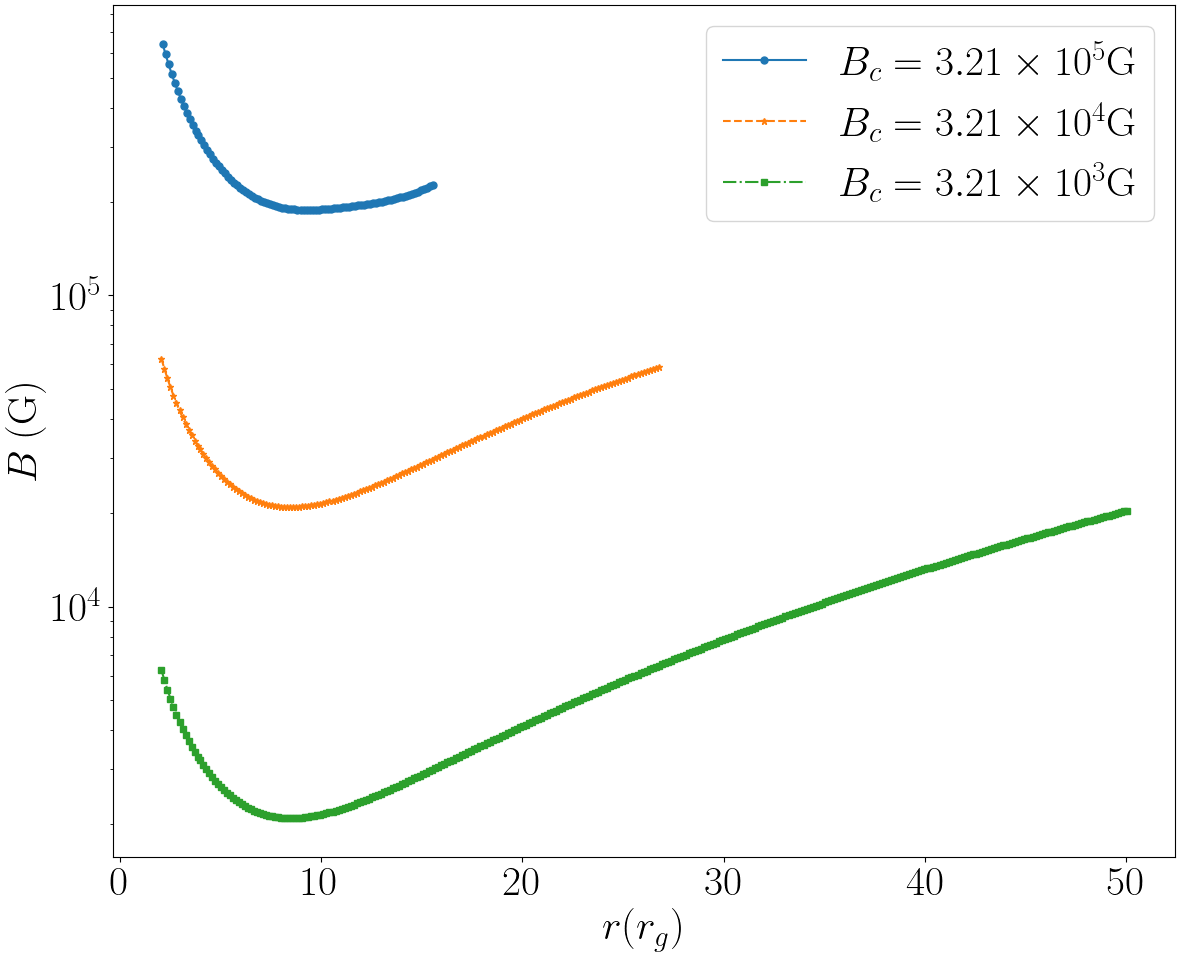}\label{a9-b}
}
     \caption{Variation of overall spectra and magnetic field profiles with the critical point magnetic field $B_c$ for $M=10M_\odot$. The other parameters are same as in Fig.~\ref{spec1}. The first, second and third peaks in the spectra are from synchrotron radiation, SSC and EC respectively.}
     \label{spec-b}
 \end{figure*}

\begin{figure*}
\centering
\subfloat[]{
\includegraphics[width=0.48\textwidth]{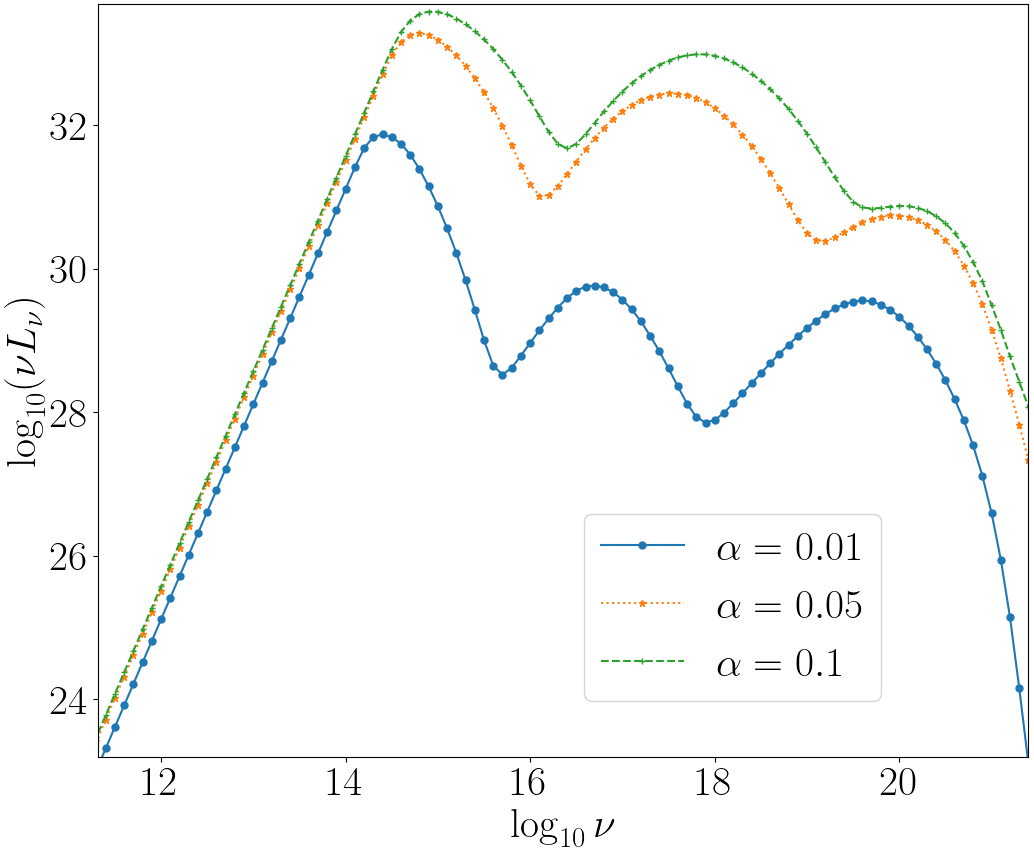}\label{s-alpha}}
\subfloat[]{
\includegraphics[width=0.49\textwidth]{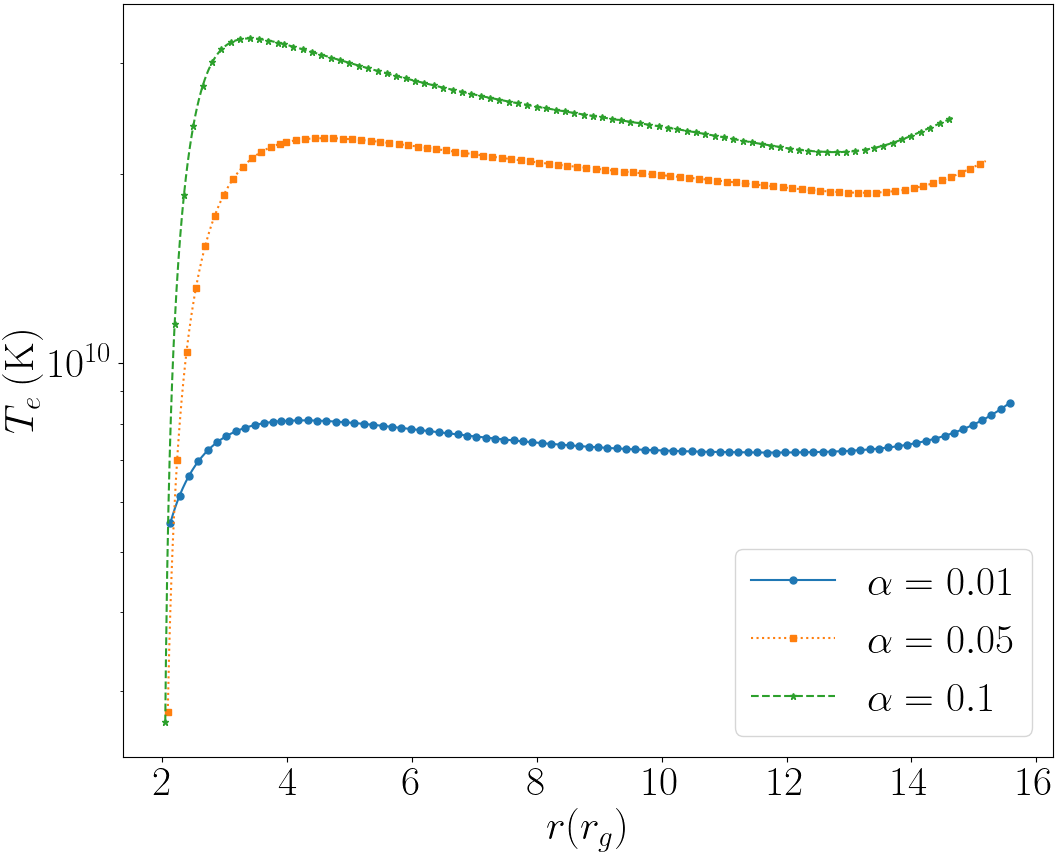}\label{alpha-te}
}
     \caption{Variation of overall spectra and $T_e$ profiles with $\alpha$ for $M=10M_\odot$. The other parameters are same as in Fig.~\ref{spec1}}
     \label{spec-alp}
 \end{figure*}

 \begin{figure*}
\centering
\subfloat[]{
\includegraphics[width=0.48\textwidth]{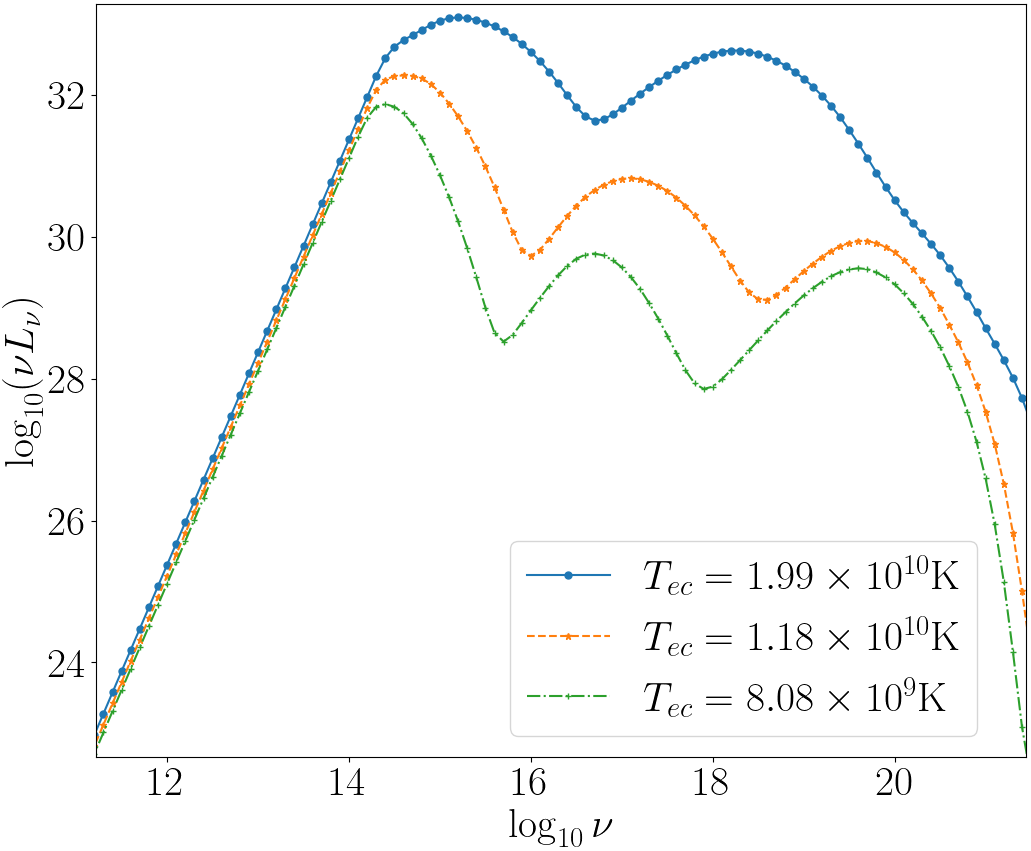}\label{s-tec}}
\subfloat[]{
\includegraphics[width=0.49\textwidth]{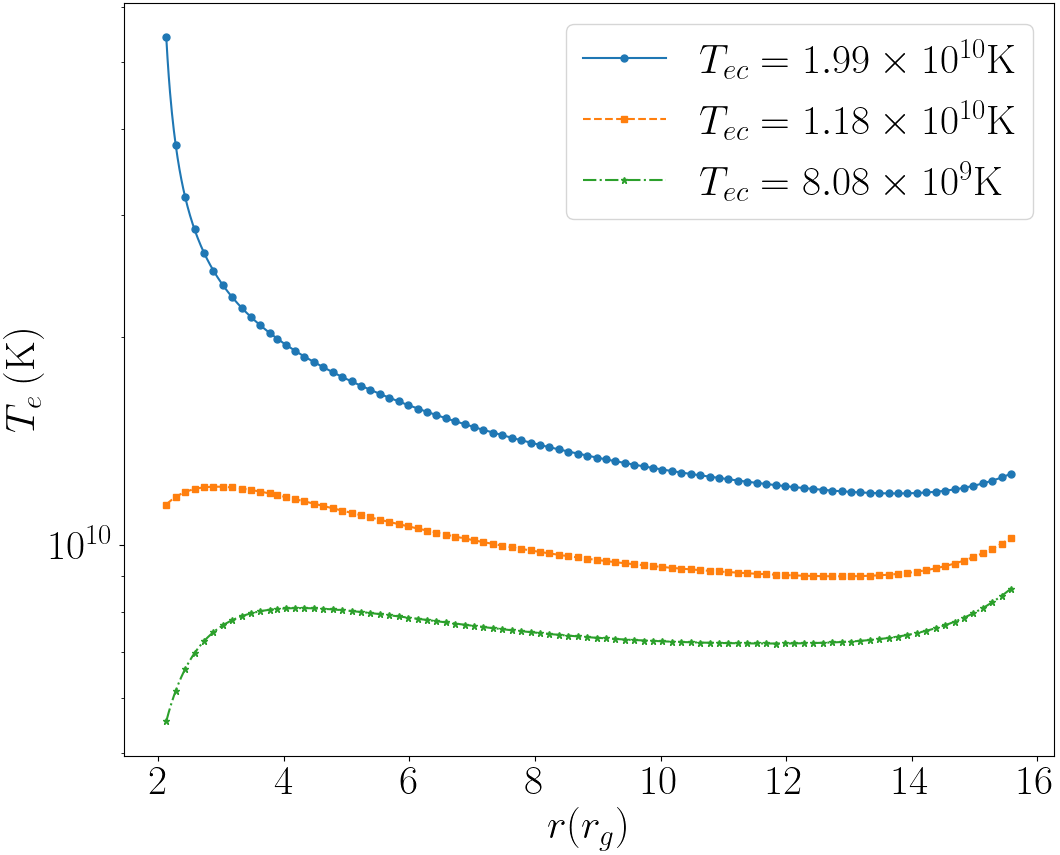}\label{a9-tec}
}
     \caption{Variation of overall spectra and $T_e$ profiles with critical point electron temperature $T_{ec}$ for $M=10M_\odot$. The other parameters are same as in Fig.~\ref{spec1}}
     \label{spec-tec}
 \end{figure*}

  \begin{figure*}
\centering
\subfloat[]{
\includegraphics[width=0.48\textwidth]{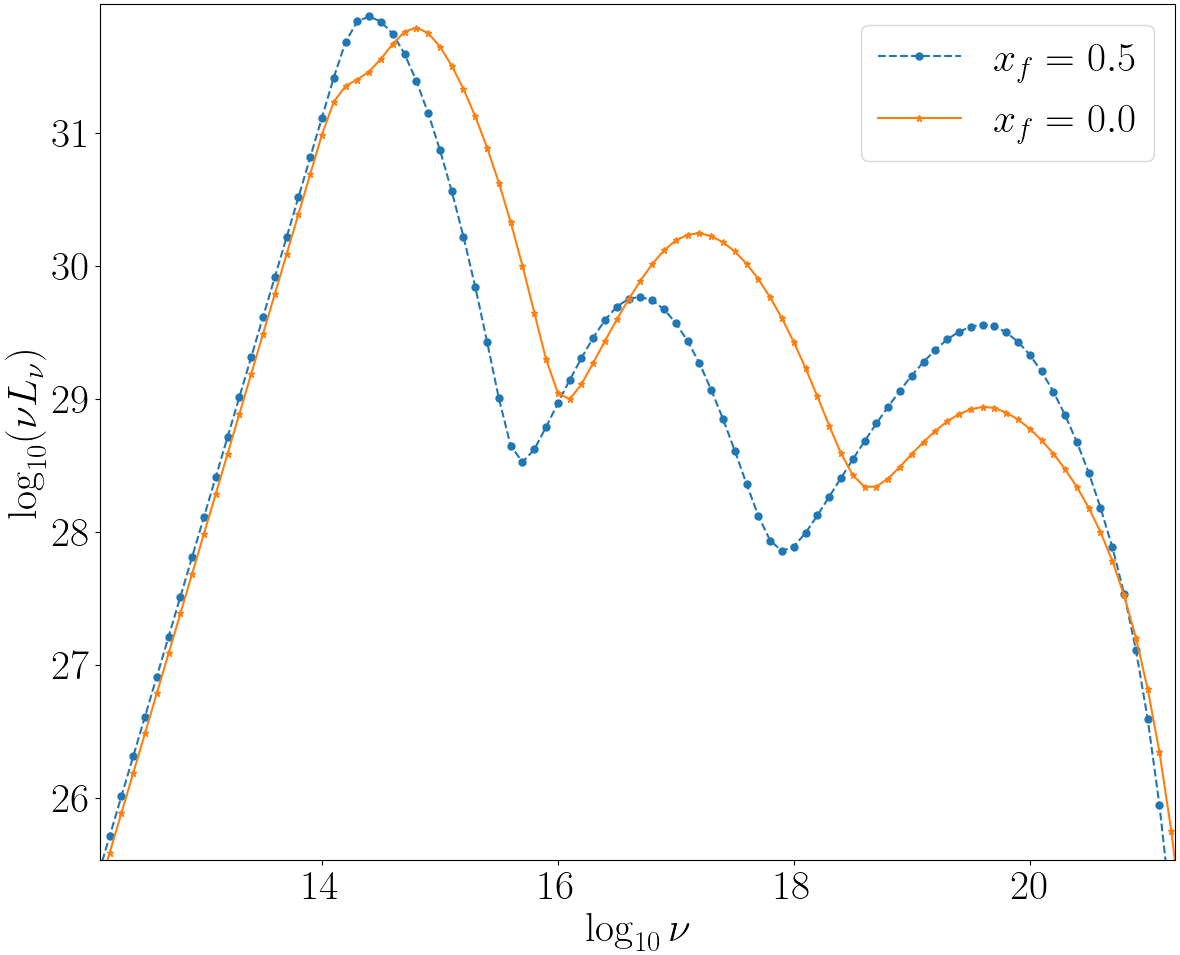}\label{s-xf}}
\subfloat[]{
\includegraphics[width=0.47\textwidth]{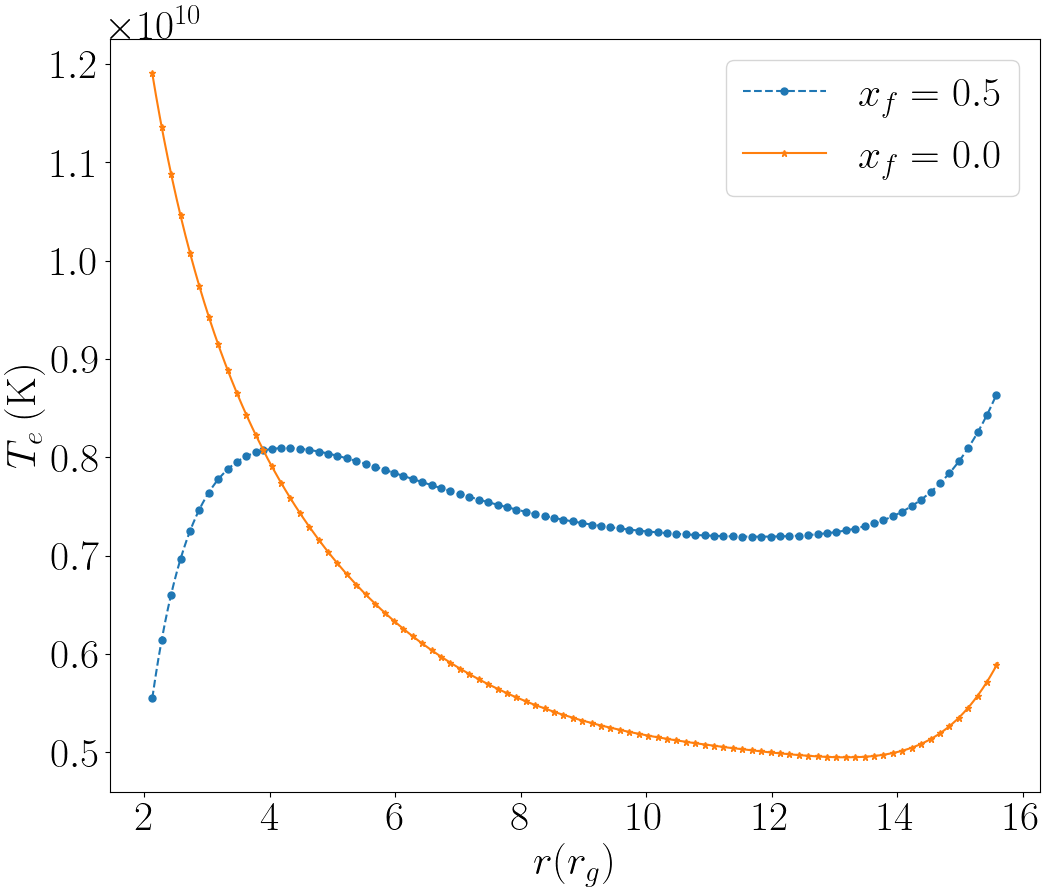}\label{a9-xf}
}
     \caption{Variation of overall spectra and $T_e$ profiles with $x_f$ for $M=10M_\odot$. The other parameters are same as in Fig.~\ref{spec1}}
     \label{spec-xf}
 \end{figure*}

 \begin{figure*}
\centering
\subfloat[]{
\includegraphics[width=0.48\textwidth]{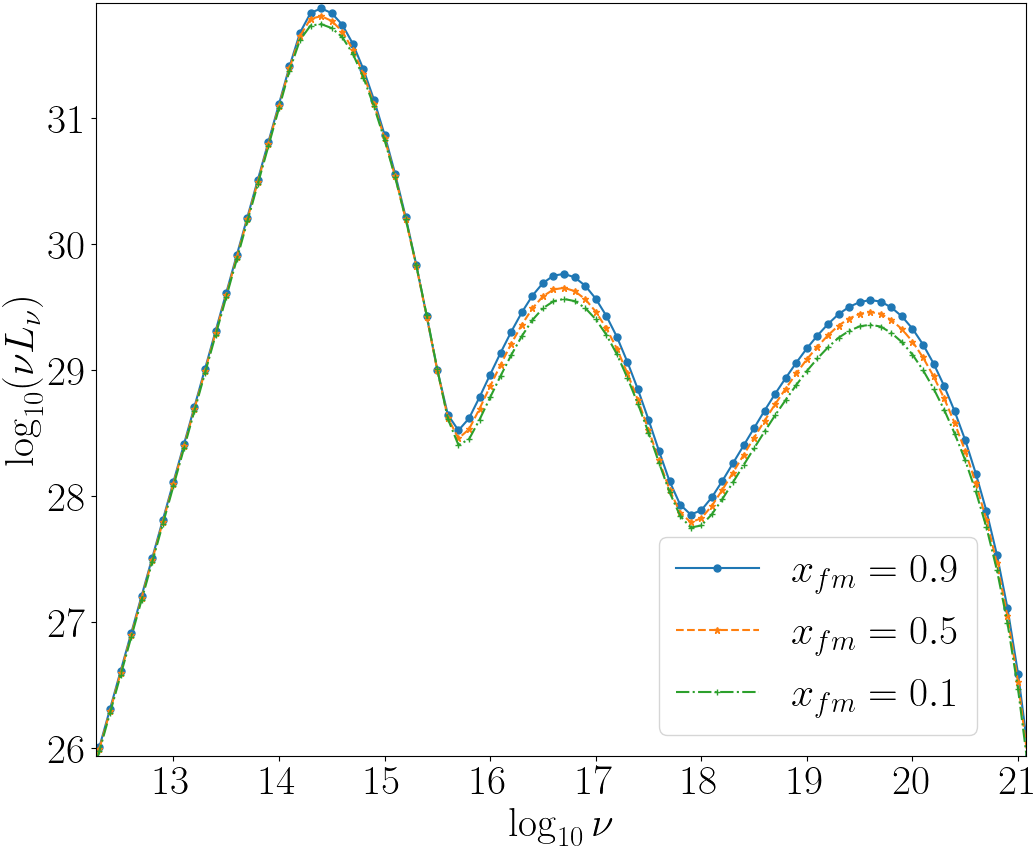}\label{s-xfm}}
\subfloat[]{
\includegraphics[width=0.485\textwidth]{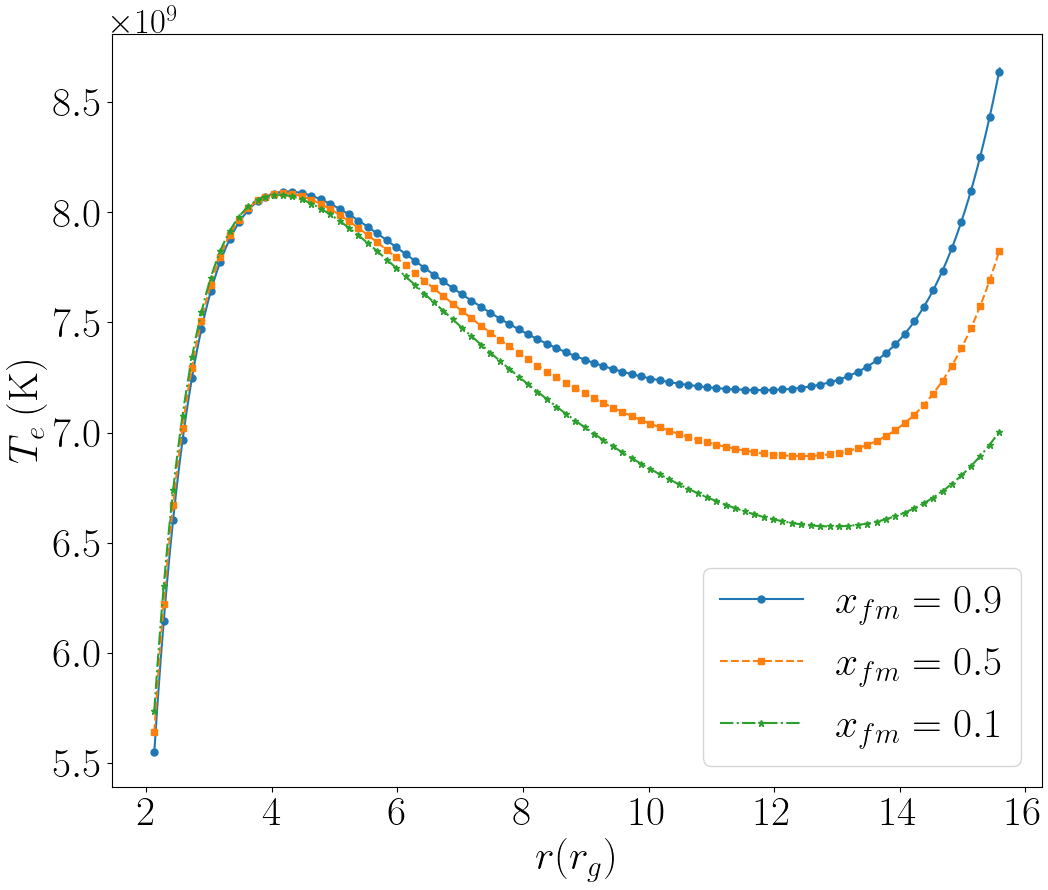}\label{a9-xfm}
}
     \caption{Variation of overall spectra and $T_e$ profiles with $x_{fm}$ for $M=10M_\odot$. The other parameters are same as in Fig.~\ref{spec1}}
     \label{spec-xfm}
 \end{figure*}

The variation of the spectrum with the spin ($a$) of a SBH is shown in Fig.~\ref{spec-a}. The synchrotron radiation and SSC peaks reduce in size and also shift to the left as $a$ decreases. This is a consequence of lower magnetic field in low spinning systems, and the fact that for high spinning black holes, the photons originate from closer to the black hole event horizon, leading to it being of higher energy. The EC peak for $a=0$ case appears very closer to the contribution from bremsstrahlung, leading to a more broader corresponding profile, as evident in Fig~\ref{spec-a}. The reduction of luminosity with spin has also been obtained using GRMHD simulations \citep{rn2011,lasota,rn2022,mp-bm}.

We next show the variation of spectra with the accretion rate of the flow in Fig~\ref{s-mdot}. The peaks of overall spectrum reduce as $\dot{M}$ is decreased. This is because of reduction in the density of the flow with decreasing $\dot{M}$. This behavior is shown in Fig~\ref{a9-den}. A reduced density leads to fewer electrons producing each component of the spectrum. In addition to the density, magnetic field is also reduced with decreasing $\dot{M}$. All of these factors combine into the decreasing trend of the overall spectral profile observed in Fig.~\ref{spec-mdot}.

As synchrotron power has a strong dependence on  magnetic field strength, we study the variation of the spectra with different magnetic field strengths. In our numerical MHD disk solutions, the angular momentum gets transported by magnetic shear ($B_iB_j$), in addition to the $\alpha$-viscosity prescription (see Eq.~\ref{ang-mom}). The magnetic field strength and profiles are assigned at the critical point ($B_c$) and, accordingly, spatial evolution of the magnetic field sets in. We consider magnetic field variations in the range of $10^3-10^5$G, which are typical field strengths observed in accretion flows \citep{zd-2014,2015ASPC..494..114P}. The magnitude and radial profile of the magnetic field depend on $B_c$ (Fig.~\ref{a9-b}). The increase of $B_{c}$ leads to an overall higher magnetic field which in turn leads to alterations in the angular momentum transport in the disk (Eq.~\ref{ang-mom}). Thus, changing $B_c$ alters the angular momentum transport due to the change of magnetic shear and also leads to a difference in disk size as shown in Fig~\ref{a9-b}. The synchrotron radiation and SSC peaks in Fig.~\ref{s-b} move to the left as the energy of the synchrotron photons reduces with decreasing magnetic field. Interestingly, the luminosity of SSC photons increases for $B_c=3.21\times10^3\text{G}$ compared to that of $B_c=3.21\times10^4\text{G}$ case. This is because the disk size almost doubles for the former than the latter (see Table~\ref{params}), leading to a larger amount of Compotonization of synchrotron photons. Due to the effect of the varying disk sizes, the EC luminosity also changes. For systems with larger disk sizes (lower magnetic fields), the amount of EC scattering increases as the incoming soft photons from the Keplerian region encounter more sub-Keplerian disk plasma, leading to higher peak EC power. This leads to the interesting effect that the EC peak behaves in an opposite manner to the synchrotron peak, i.e. it increases with decreasing $B$, while the latter decreases.

The temperature of the accretion flow also plays a significant role in the characteristics of the spectrum. We study the effects of variations in electron heating and electron temperature, assigned at the critical point, on the spectral characteristics. 
In Fig~\ref{alpha-te}, we show the $T_e$ profiles for different values of $\alpha$. Although $\alpha$-viscosity, in general, leads to angular momentum transport, in the present work the angular momentum is mainly transported by the strong magnetic fields  and, thus, our disk sizes are almost same for three $\alpha$ considered.

As the electrons in our system get heating from viscous and Ohmic heating in addition to Coulomb coupling with protons, the $\alpha$-viscosity parameter strongly affects $T_e$. This leads to corresponding changes in the spectrum as shown in Fig.~\ref{s-alpha}. As $T_e$ increases with $\alpha$, the peaks of the spectra shift to the right and also with the increase in luminosity. The SSC peak becomes broader and the ratio of synchrotron to SSC luminosity reduces, indicating significant increase in SSC luminosity with the increase of $\alpha$.

Variation of $T_e$ defined at the critical point, namely  $T_{ec}$, leads to the changes in the temperature profile throughout the disk as shown in Fig.~\ref{a9-tec}. We have chosen three $T_{ec}$ values such that $T_e$ shows concave to almost flat to convex characteristics as $T_{ec}$ increases. These $T_{ec}$'s also lead to $T_e$'s in the range of $10^9-10^{10}\text{K}$, which are typical electron temperatures for advective accretion flows \citep{tcaf,adaf,rajesh10}. The interplay between electron heating via Coulomb coupling, gravity and cooling leads to different $T_e$ profile shapes. The corresponding spectra are shown in Fig.~\ref{s-tec}. For the highest $T_{ec}$, the spectrum has only two prominent peaks generated by synchrotron and SSC emissions. This is caused by the increase in the width of the SSC peak with increasing $T_e$. The peak thus ends up enveloping the EC peak, which is only visible as a tail after $\nu=10^{20}\text{Hz}$. For the other two $T_{ec}$ values, we see similar behavior as with $\alpha$ variation. The peaks shift to the right, the luminosity increases and the SSC peak broadens with increasing $T_e$. The spectra is very sensitive to $T_e$ and its profile. Upon variation of $T_{ec}$ by $\sim1.5$ times (Fig.~\ref{a9-tec}), the bolometric luminosity increases by an order of magnitude and the shapes of the various peaks are strongly affected as well, most notably the SSC peak.

The amount of heat received by the electrons via the viscous and Ohmic heating mechanisms is given by $x_f$ and $x_{fm}$ parameters. 
In Fig.~\ref{spec-xf}, we show the variation of spectra and $T_e$ profiles with $x_f$. As $x_f$ increases, $T_e$ increases in the disk region outside the critical point. This trend reverses in the region between the inner edge and the critical point (see Fig.~\ref{a9-xf}). As $x_f$ decreases, the heat transferred to the electrons reduces leading to lower temperatures outside the critical point. This also lowers the electron cooling, leading to an increase in $T_e$ close to the black hole. This leads to non-trivial effects on the spectrum of the flow. For $x_f=0$, we observe a shift to the right in the synchrotron radiation and SSC peaks. Although the synchrotron peak is lower, the SSC peak for this case is larger than the $x_f=0.5$ case. This is because of the larger temperature near the black hole for $x_f=0$ case. As we have observed in previous cases, the SSC peak is very sensitive to the $T_e$ characteristics of the flow. This sensitivity leads to the non-trivial behavior of the $x_f=0$ case. Despite this, the variation on the overall luminosity is small with the ratio of the bolometric luminosity for the $x_f=0.5$ and $x_f=0.0$ cases being 1.21. 

Similarly, we also obtain the electron temperature profiles with the variation of the Ohmic heating  fraction $x_{fm}$ received by the electrons. The $T_e$ profiles are shown in Fig.~\ref{a9-xfm}. Contrary to the $x_f$ variation, the effect of $x_{fm}$ is much smaller on $T_e$. A small reduction in $T_e$ is obtained at the other edge of the disk as $x_{fm}$ reduce. Thus, the corresponding spectra shown in Fig.~\ref{s-xfm} also show small variations with the ratio of the bolometric luminosity for the $x_{fm}=0.9$ and $x_{fm}=0.1$ cases being 1.33. The plasma-$\beta$ (gas pressure/magnetic pressure) of our accretion flow is larger than unity throughout ($2<\beta_m<13$). This makes the viscous heating more important in the context of $T_e$, leading to higher non-triviality in the spectra for $x_f$ variations. 


\subsubsection{Luminosity and disk size}

The maximum bolometric luminosity ($L$) obtained from our parametric variations corresponding to stellar mass black hole of typical $\sim 10 M_{\odot}$ is $\sim10^{33}~\text{erg/s}$, which corresponds to a radiative efficiency ($L/\dot{M}c^2$) of around $10^{-3}$. Such radiative efficiency usually corresponds to the black hole X-ray binaries (BHXRBs) in low-hard state (LHS) or their possible AGN counterparts: 
low excitation (LE)/low luminous (LL) FRI-type radio galaxies (RGs) (FRI-LERGs/FRI-LLAGNs) (e.g., \citealt{merloni-2003}; \citealt{kording-2006}; \citealt{juan-2021}; 
\citealt{ghosh-jheap}, and references therein). Here it needs to be noted that in our analysis we have considered 1.5-dimensional accretion flow with no vertical velocity ($v_z$). However, in realistic accretion scenarios around black holes, usually advective accretion flows are associated with outflows and jets. If they are strongly advective or advection dominated flows, they are typically associated with low-power continuous, steady radio jets as observed in LHS BHXRBs and in FRI-LERGs/FRI-LLAGNs. If they exhibit inner moderately advective flows, then they are supposedly associated with more powerful relativistic jets, typically observed in transition state of BHXRBs (hard-intermediate state during LHS$-$to$-$high soft state (HSS) transition) or in their possible broad counterparts in radio loud (RL) quasars or high excitation RGs (HERGs) 
(e.g., \citealt{liu-2007}, \citealt{heckman-2014}; \citealt{zhu-2020}; \citealt{juan-2021}; \citealt{moravec-20222}; \citealt{ghosh-jheap}, and references therein). The presence of outflows (with $v_z$ component) can lead to losing $B$ along with matter and hence a decreasing radial profile for $B$ is possible. This will allow for bigger sub-Keplerian accretion disks (or the truncation of the Keplerian disk further away) as the angular momentum transport by the magnetic shear will decrease away from the black hole as shown by \cite{ULX20}. This decreasing $B$ profile with larger sub-Keplerian disk will also eventually allow for stronger $B$ close to the black hole, leading to higher synchrotron and bolometric luminosity. Such an accretion disk configuration may also be affected by $x_f$ and $x_{fm}$, as the flow would be hotter and more magnetized \citep{ghosh-jheap}, particularly in the inner region, leading to higher luminosity and with much higher radiative efficiency. This scenario is more relevant in the context of bright (hard$-$) intermediate state in BHXRBs or RL quasars and HERG in AGNs. In essence, this quite resembles hot luminous advective flows or radiatively efficient hot flows \citep{xie-2012}, with interesting observational consequences. Such a study with the incorporation of outflows is left for future work. 

\section{Spectra from GRMHD simulations} \label{grmhd}

In the preceding section, using our numerical steady state solution for 1.5-dimensional accretion flow, we have studied the effect of various disk parameters on the spectrum. These parameters affect the disk size, magnetic field, density and electron temperature of the accretion system. We now study the effects of magnetic field configuration/evolution on the spectrum using GRMHD simulations. 

\subsection{Simulation set-up}\label{setup}

We explore the spatio-temporal evolution of the magnetized accretion flow around a rotating black hole using the publicly available GRMHD code Black Hole Accretion Code (\texttt{BHAC}) \citep{bhac}. The accretion flow is initiated by using the Fishbone-Moncrief (FM) \citep{fm} torus solution. 

The equations solved by the codes are:
\begin{equation}
\begin{aligned}
    \nabla_\mu(\rho u^\mu)&=0,\\
    \nabla_\mu T^{\mu\nu}&=0,\\
    \nabla_{\mu}\hspace{0.01in}^*F^{\mu\nu}&=0,   
\end{aligned}
\end{equation}
where $u^{\mu}$ is the four-velocity, $T^{\mu\nu}$ is the stress-energy tensor and $^{*}F^{\mu\nu}$ is the dual Faraday tensor. Here, $\mu$ and $\nu$ are spacetime indices such as $t,r,\theta,\phi$.

We carry out 2.5-dimensional simulations, by exploiting the axisymmetry of the system. Our analysis is based on time and density averaged profiles of flow variables like density, magnetic fields, velocity etc. It is known that 2.5-D GRMHD simulations show quite distinct characteristics than 3-D simulations. Most notably the flux-eruption events, characteristic of MAD simulations, are quite exaggerated in 2.5-D as compared to 3-D. This leads to a difference in their temporal characteristics. However, for the present purpose of analyzing accretion disk spectra, corroborated with observations, time-averaged quantities suffice. Hence, we do not expect much difference in the flow properties of our interest between 2.5-D and 3-D simulations. The computational inexpensiveness of 2.5-D simulations also allows us in extending our spectral analysis to multiple black hole spins and magnetic vector potentials.

The simulations have been run at a resolution of $384\times192\times1$. All simulations are evolved to $3\times10^4$ timesteps. This resolution results in effectively resolving the ergosphere and the regions very close to the event horizon, where most of the GR effects of the accretion flow take place. This is important as magnetic fields and outflows lead to profound effects on the dynamics of the accretion flow. The chosen duration of the simulation allows the disk size to be around 14 $r_g$ (inflow-outflow equilibrium radius; see below for details). This leads to disk sizes similar to those analyzed in the our numerical steady state calculations. As shown in Sec.~\ref{para}, the disk size is a very important factor in the calculations of spectra. Considering similar disk sizes lead to effective comparisons between the steady state and GRMHD simulation calculations.
To study the effects of black hole spin on the outflow power, we consider two black hole spins in our simulations, $a=Jc/(GM^2)=$ 0.5 and 0.9375.

The magnetic field evolution is initiated by two formalisms, namely SANE and MAD. Their corresponding magnetic vector potentials are given by \citep{fluxerr,mp-bm}:
\begin{enumerate}
    \item SANE: $A_{\phi}=\max(\rho/\rho_0-0.2,0)$,
    \item MAD: $A_\phi=\exp(-r/400)(r/r_{\mathrm{in}})^3\sin^3\theta\max(\rho/\rho_{0}-0.01,0)$,
\end{enumerate}
where $\rho_0$ is the maximum density in the initial FM torus, set at $r=41r_g$, while $r_{\mathrm{in}}=20r_g$ is the inner edge of the FM torus.
The initial field strength is set up by defining the initial plasma-$\beta$ parameter to $100$, where magnetic pressure is $b^2/2$ with $b^2=b_\mu b^\mu$ and $b^\mu$ being the four-magnetic field. Starting with a high plasma-$\beta$ allows for the accretion to progress effectively. If we start with a low plasma-$\beta$, the magnetic field will build up quickly. This results in block off of the accretion from reaching the black hole, even blowing up of the initial torus, due to high magnetic pressure. A much higher plasma-$\beta$ ($\gg100$), on the other hand, will lead to the accretion progressing very slowly as the angular momentum transport due to the resultant MRI would be very weak, thus requiring long simulation durations.

Build up of large magnetic fields in a given volume leads to numerical errors in GRMHD codes \citep{ress}. To curb this, we have set the maximum value of the magnetization ($\sigma=b^2/\rho$) to be $\sigma_{max}=100$. If this limit exceeds, matter density is injected in the coordinate frame to bring $\sigma$ below 100.

If the magnetization is not controlled as mentioned above, due to advection and flux freezing, magnetic flux will continue to accumulate near the black hole. This will lead to the formation of a magnetic barrier due to excess magnetic pressure, blocking off the incoming flow. As soon as $\sigma$ becomes more than a threshold value ($\sigma_{max}=100$ here), the density floors are reset in accordance with the threshold $\sigma$ value. This in turn increases the plasma-$\beta$, allowing for inflow to resume.

\subsection{Disk properties}

The accretion flow in our simulations undergoes inflow and outflow at each radius throughout the evolution of the simulation domain. Given the large size of our initial matter torus, any out flown matter is replenished by inflow. This eventually leads to an inflow-outflow equilibrium out to a certain radius, subject to the evolution time and the time averaging window \citep{fluxerr}. The matter injection prescription based on $\sigma$ also affects the inflow-outflow equilibrium radius. If the threshold value of $\sigma$ ($\sigma_{max}$) is set to a low value, accretion rate would increase unphysically, as too much matter will be injected as soon as the field exceeds some lower value. This will also render the magnetic field dynamically less important, as $b^2/\rho$ will remain low throughout the flow. If $\sigma_{max}$ is set to a high value, the magnetic barrier formed due to high magnetic pressure will persist for a longer simulation time, leading to again unphysical results. We do not expect, however, much difference in our results if small variations from the chosen $\sigma_{max}$ are made. This is because our spectral analysis relies on time and density averaged flow variables.

\begin{figure*}
\centering
\subfloat[SANE]{
\includegraphics[width=0.49\textwidth]{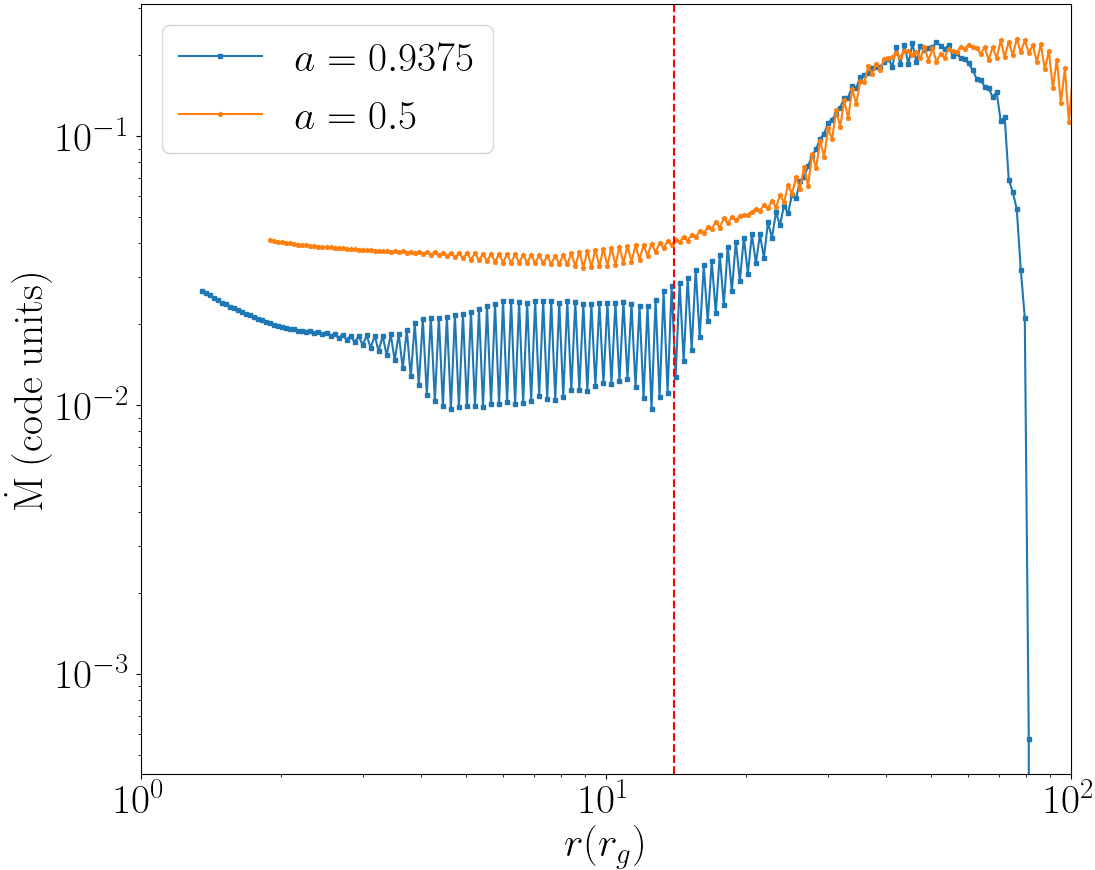}\label{sane-mdot}}
\subfloat[MAD]{
\includegraphics[width=0.49\textwidth]{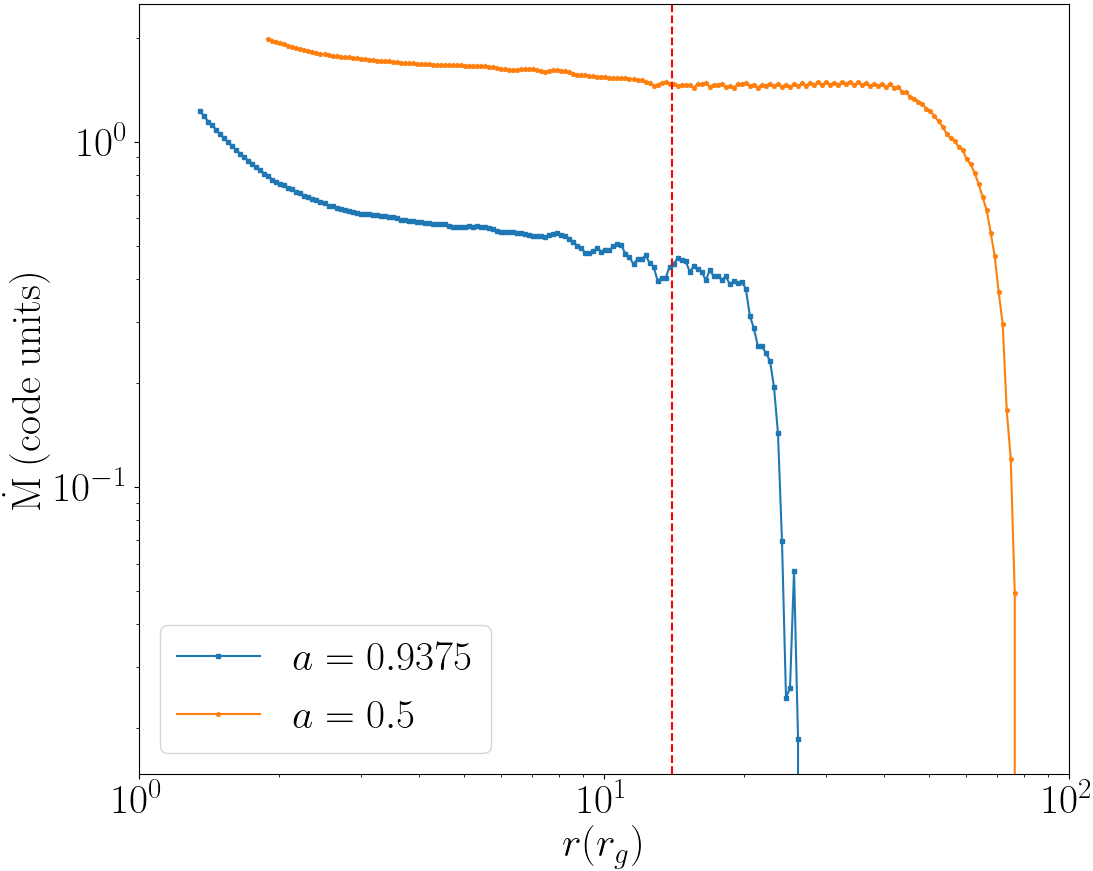}\label{mad-mdot}
}
     \caption{Time averaged radial accretion rate profiles for SANE and MAD simulations with the inflow-outflow equilibrium boundary marked by the red dashed line.}
     \label{mdot}
 \end{figure*}

To obtain the inflow-outflow equilibrium in our simulations, we calculate the accretion rate of the flow, defined by
\begin{equation}
    \dot{M}(r)=-\int\sqrt{-g}\rho u^r\mathrm{d}\theta \mathrm{d}\phi,
\end{equation}
where $\dot{M}(r)$ is time averaged over 15000-30000 $r_g/c$ time-steps and the obtained accretion rate is shown in Fig.~\ref{mdot}. The radial extent of the inflow-outflow equilibrium region depends on the time-averaging window. We choose the aforementioned timestep window to allow the inflow-outflow equilibrium radius to extend out to $14r_g$, which leads to a disk size similar to our steady state numerical calculations. This enables effective comparisons between our simulation and numerical results.

In Fig.~\ref{mdot}, the red dashed vertical line at $r=14r_g$ indicates the inflow-outflow equilibrium radius in our simulations. The SANE simulations go out of equilibrium after this radius. Although the MAD $\dot{M}$ profiles are in inflow-outflow equilibrium till larger radii, to compare our MAD and SANE results we restrict our disk to $r=14r_g$.

We emphasis that the black hole accretion is necessarily transonic, as shown in our steady state calculations (Fig.~\ref{vel}) (see also, \citealt{rajesh10,bm-kc,ULX20}). Hence, we calculate the quantities relevant for spectra in the locally non-rotating frame (LNRF). This enables us to study the simulation system in a frame in which the accretion flow is transonic. 

The four velocity and four magnetic fields obtained from \texttt{BHAC} in Boyer-Lindquist coordinates are transformed in LNRF using tetrads as $u^i=\Lambda^i_\nu \tilde{u}^\nu$, where $\tilde{u}^\nu$ is any four-vector in the fluid frame, $u^i$ is the transformed four-vector in LNRF and $\Lambda^i_\nu$ is the transformation matrix containing tetrads defined by \cite{bl-tet}.

The obtained four velocity in LNRF is transformed to three velocity by the transformation: $v^i=u^i/u^t$, where $i\in\{x,y,z\}$.

The four magnetic fields in LNRF are transformed to three magnetic fields by the transformation: 

\begin{equation}
    B^i=b^iu^t-b^tu^i
\end{equation}
where $i\in\{x,y,z\}$.

These quantities ($q$) are calculated over one disk scale height defined by:
\begin{equation}
    \frac{h}{r}=\frac{\int\sqrt{-g}\rho |\theta-\pi/2|\mathrm{d}\theta \mathrm{d}\phi}{\int\sqrt{-g}\rho \mathrm{d}\theta \mathrm{d}\phi},
\end{equation}
and then density averaged using,
\begin{equation}
    <q>_{\mathrm{disk}}=\left.\frac{\int q\sqrt{-g}\rho  \mathrm{d}\theta \mathrm{d}\phi}{\int\sqrt{-g}\rho \mathrm{d}\theta \mathrm{d}\phi}\right\rvert_{h/r}.
\end{equation}

All physical quantities discussed from hereinafter are density averaged over one scale-height and time averaged over 15000-30000 $r_g/c$. We also omit the $<\cdot>_{\mathrm{disk}}$ notation for these quantities to improve readability.

\begin{figure}
\centering
\includegraphics[width=0.47\textwidth]{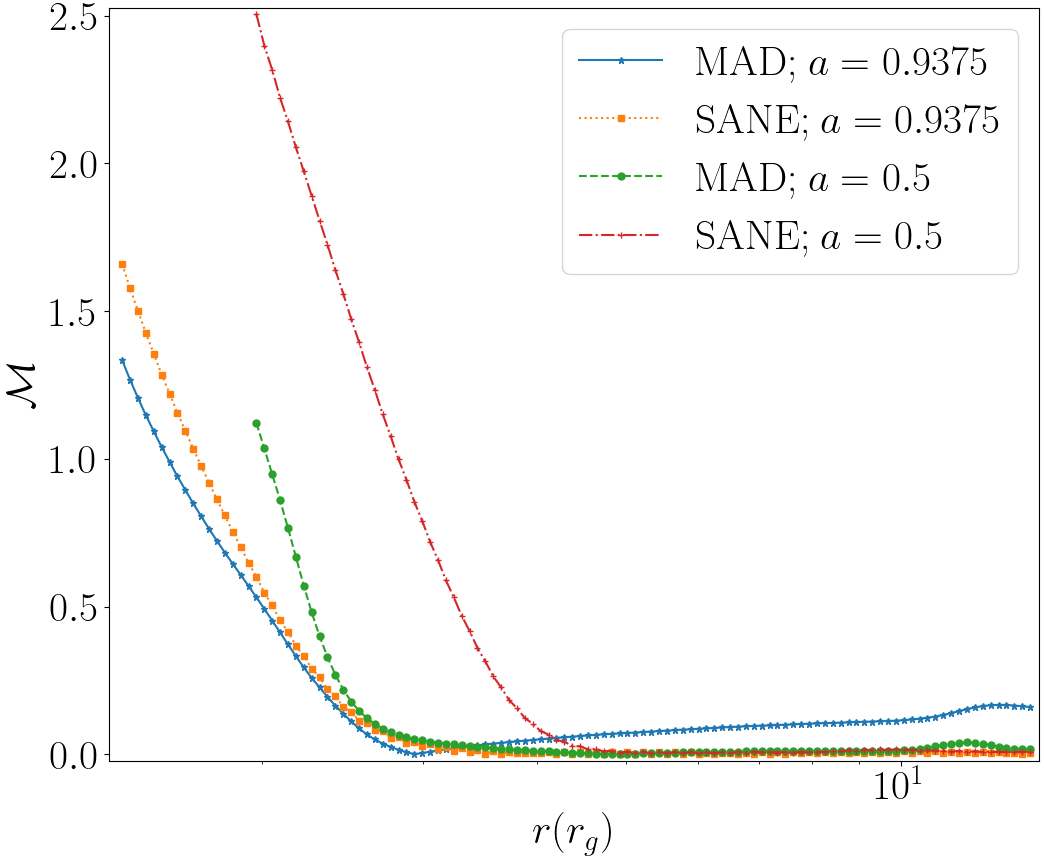}
     \caption{Radial Mach number profiles for all the simulations in the LNRF frame.}
     \label{mach}
\end{figure}

The radial Mach number ($\mathcal{M}=v^r/c_s$) profile in this frame is shown in Fig.~\ref{mach} for all the simulations. It is apparent that the flow becomes supersonic close to the black hole. As in MAD, matter accumulates close to the black hole around  a strong magnetic barrier, flows are geometrically thicker and compressed. This leads to a hotter flow. Hence, their $\mathcal{M}$ are lower and the flow becomes supersonic only very close to the event horizon of the black hole, as compared to SANEs. Without magnetic barrier and compression, SANE remains cooler with larger $\mathcal{M}$.

\begin{figure*}
\centering
\subfloat[$a=0.5$]{
\includegraphics[width=0.49\textwidth]{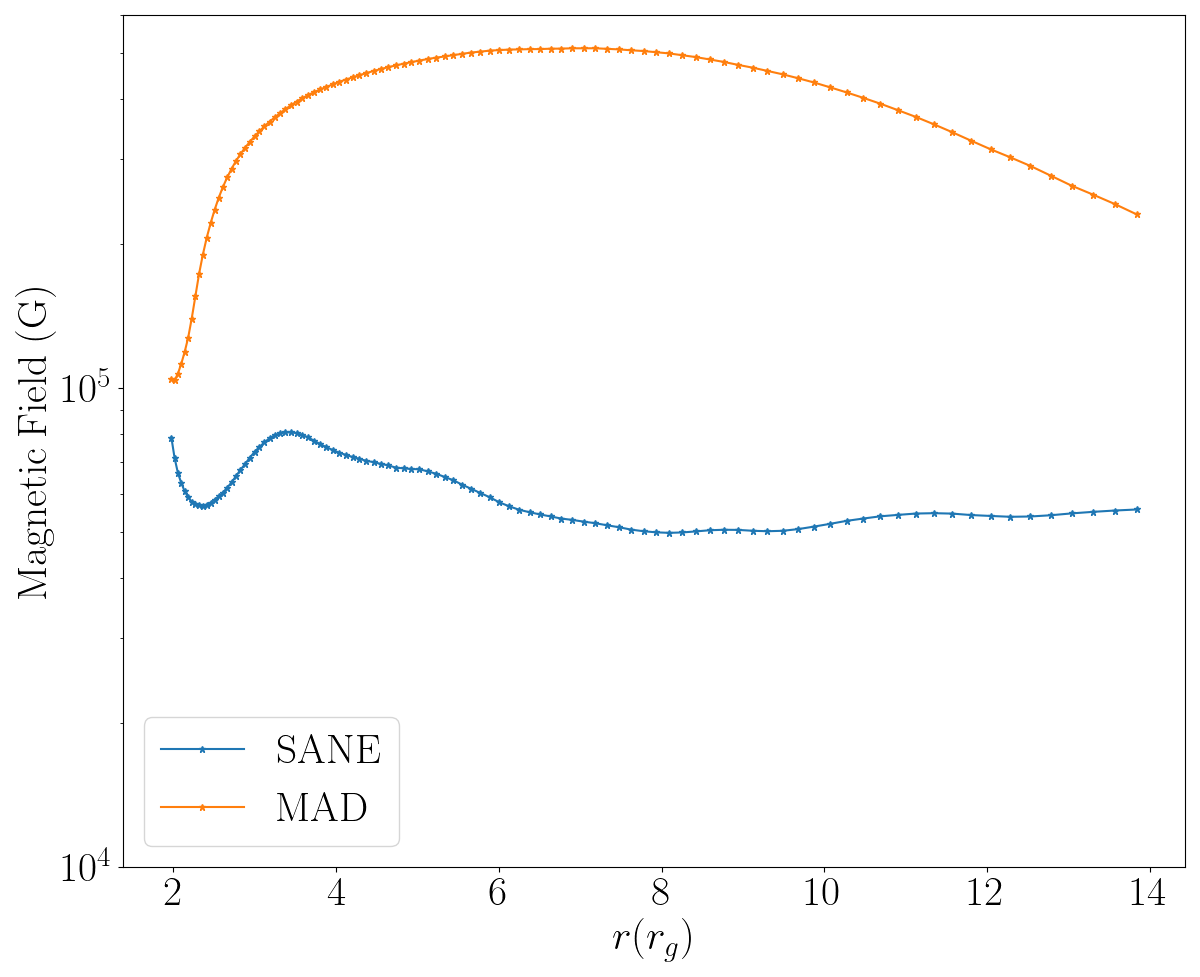}\label{mag-a5}
}
\subfloat[$a=0.9375$]{
\includegraphics[width=0.49\textwidth]{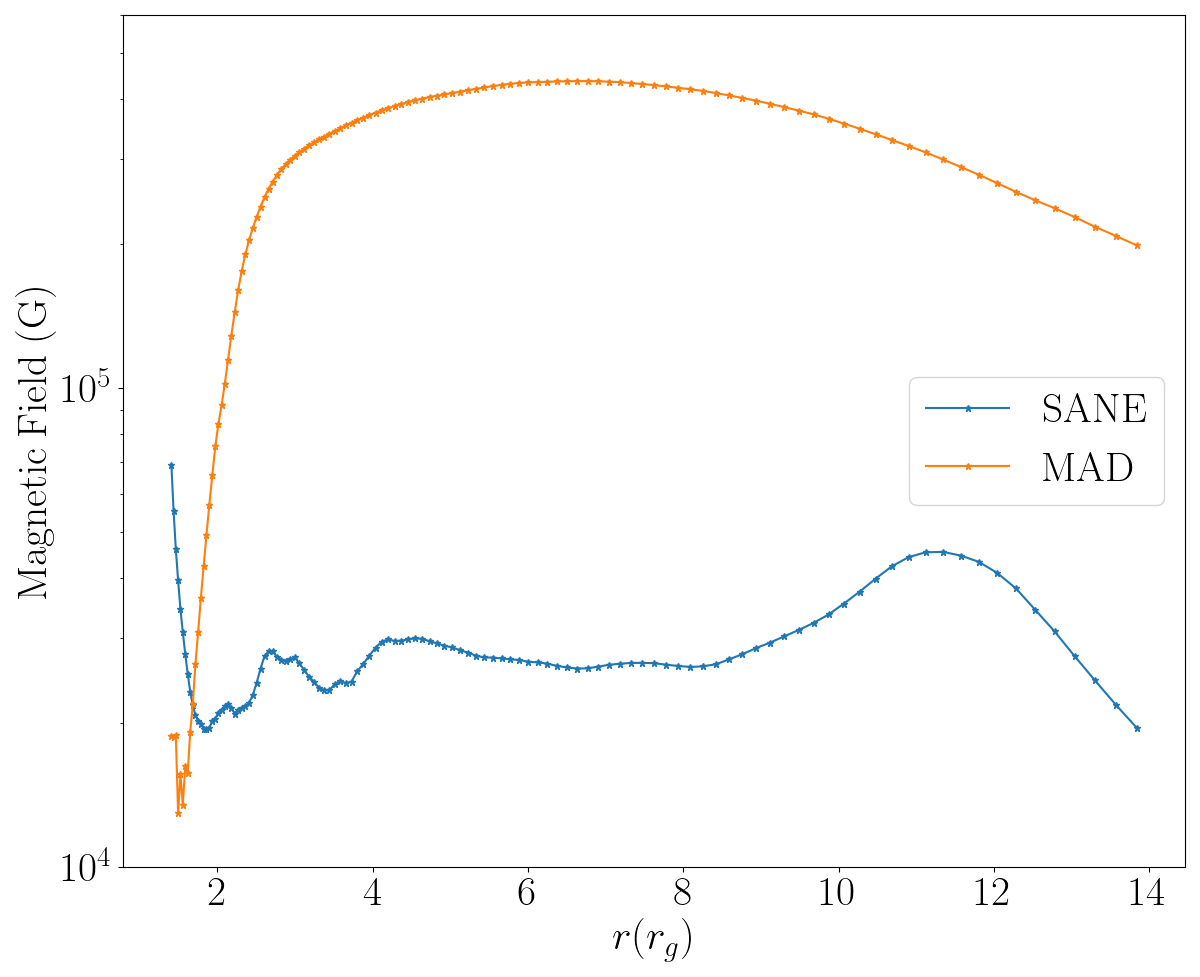}\label{mag-a9}
}

\subfloat[$a=0.5$]{
\includegraphics[width=0.49\textwidth]{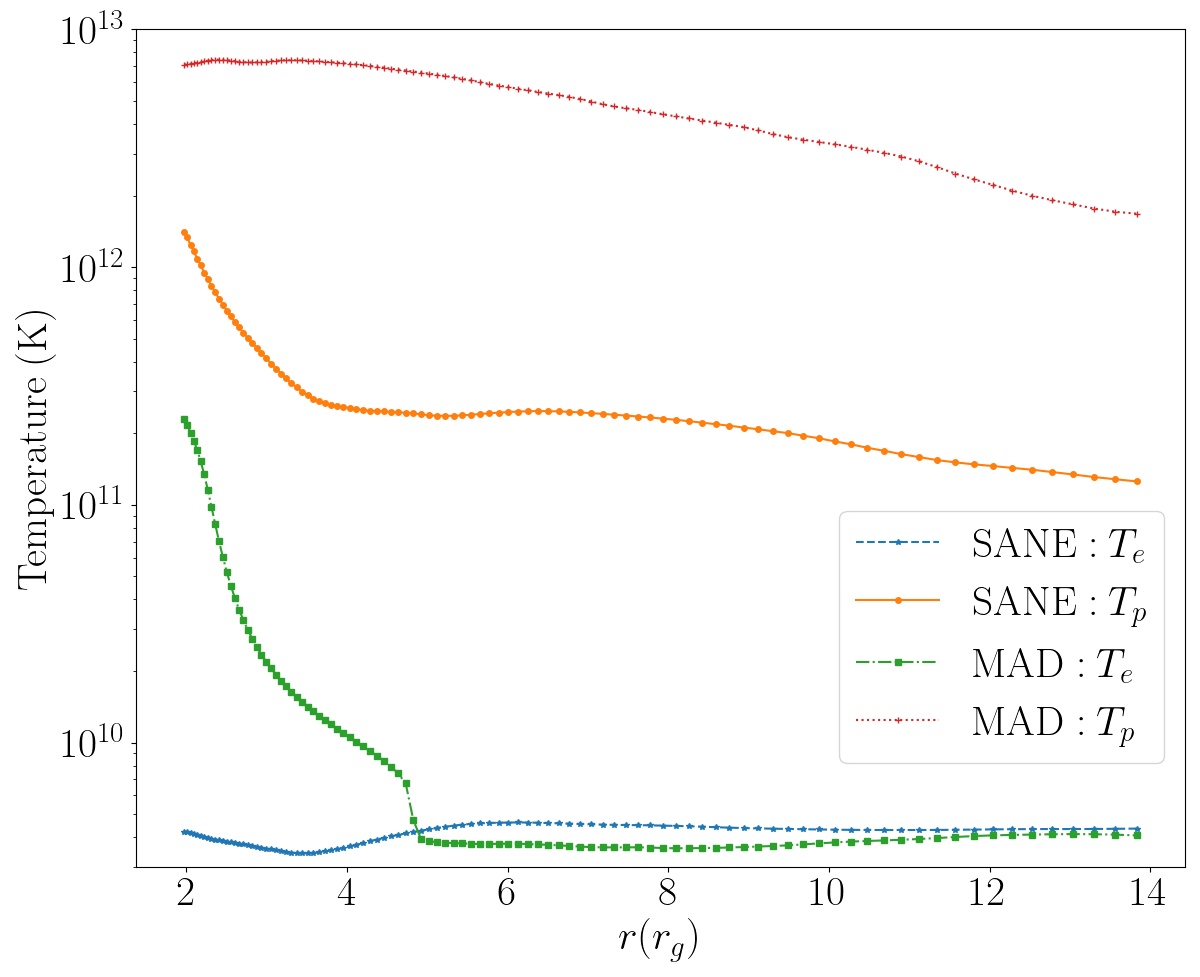}\label{temp-a5}
}
\subfloat[$a=0.9375$]{
\includegraphics[width=0.49\textwidth]{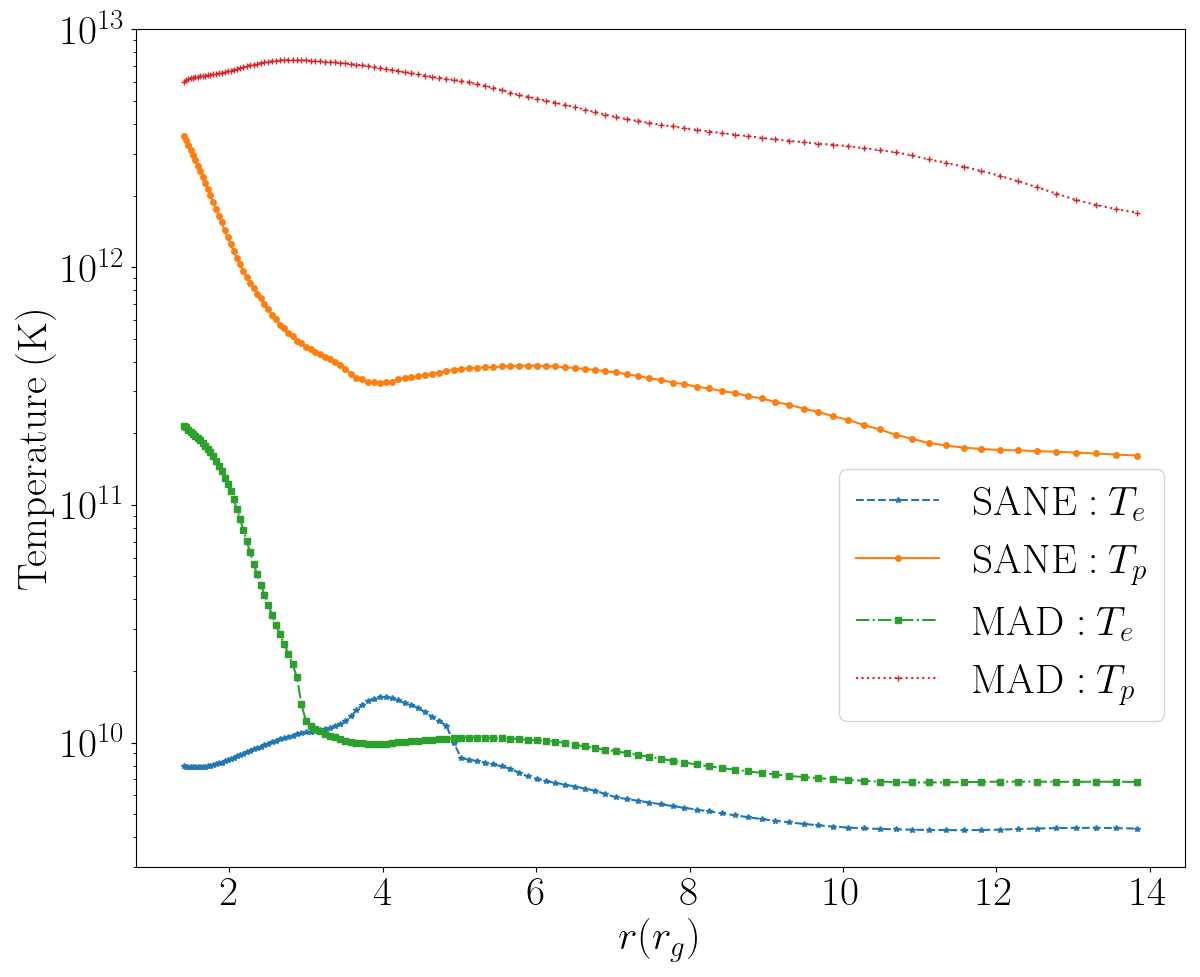}\label{temp-a9}
}
     \caption{Radial magnetic field and temperature profiles for MAD and SANE simulations for $a=0.5$ and $a=0.9375$}
     \label{sim-pro}
 \end{figure*}

The innermost stable circular orbit (ISCO) for $a=0.5$ is at $r=4.233r_g$, while for $a=0.9375$ ISCO is at $r=2.044r_g$. Below ISCO, due to strong gravity, matter falls almost radially into the black hole. For $a=0.5$, the region between the event horizon ($r=1.866r_g$) and ISCO is larger than that for $a=0.9375$, whose event horizon is at $r=1.35r_g$. This leads to flows around lower spinning black holes acquiring higher radial speeds after crossing ISCO but before approaching the horizon. The difference in the sound speeds ($\propto\sqrt{T_p}$; see Fig.~\ref{temp-a5}) for MAD and SANE for $a=0.5$ is larger than that for $a=0.9375$. This is because, due to the larger event horizon of $a=0.5$, the accretion flow terminates earlier as compared to the $a=0.9375$ case. This does not allow the sound speed difference between MAD and SANE to decrease. Fig.~\ref{temp-a9} confirms that the trend of sound speed variation in SANE for $a=0.5$ continues for $a=0.9375$ till lower radius (till its lower event horizon) decreasing the difference with MAD. This leads to the apparent difference in the behaviors of MAD and SANE $\mathcal{M}$ in Fig.~\ref{mach}.

The obtained time averaged four magnetic field profiles for SANE and MAD simulations are shown in Fig.~\ref{mag-a5} and Fig.~\ref{mag-a9}.
The MAD magnetic fields are always higher in magnitude than the SANE magnetic fields by about an order of magnitude. This is expected as MAD simulations, by definition, are highly magnetized. The magnetic field in MADs decreases close to the black hole due to the flux-eruption events, which are a characteristic feature of MAD. These eruptions take away magnetic flux with them, leading to a reduction in the magnetic field. Such events are not present in SANEs, leading to almost constant (slight increase for SANE, a=0.9375, near the black hole) magnetic field profiles.

\subsection{Electron temperature calculation}
The obtained LNRF three vectors are used in solving the electron temperature ($T_e$) equation,
\begin{dmath}
    \frac{v}{\Gamma_3-1}\left(\frac{dP_e}{dr}-\frac{\Gamma_1P_e}{\rho}\frac{d\rho}{dr}\right)=Q^{ie}-Q^{e-}, \label{sim-te}
\end{dmath}
to find $T_e$ at each radius of the flow. 
We omit viscous and Ohmic heating for electrons as they do not have significant effect on $T_e$ and subsequent bolometric luminosity, confirmed by the spectra shown in Figs.~\ref{spec-xf} and \ref{spec-xfm}. The other notations in Eq.~(\ref{sim-te}) are same as in Eq.~(\ref{sol-te}).

To solve Eq.~(\ref{sim-te}), $T_e$ value at the outer edge of the disk, i.e. $r=14r_g$, is to be provided. These values for the different simulations are presented in Table~\ref{tab:sim-temp}. 

\begin{table}
\vspace{0.2in}
\centering
\setlength{\tabcolsep}{8pt}
\begin{tabular}{c c c}
\hline
$a$ & MAD & SANE\\
\hline
0.5 & $3.2748\times10^{-4}$ & $3.5\times10^{-4}$ \\
0.9375 & $5.5017\times10^{-4}$ & $3.5\times10^{-4}$ \\
\hline
\end{tabular}
\caption{Input temperature at the outer edge of the disk for GRMHD simulations in units of $\mu_em_pc^2/k_b$.}
\label{tab:sim-temp}
\end{table}
\begin{figure*}
\centering
\subfloat[SANE]{
\includegraphics[width=0.49\textwidth]{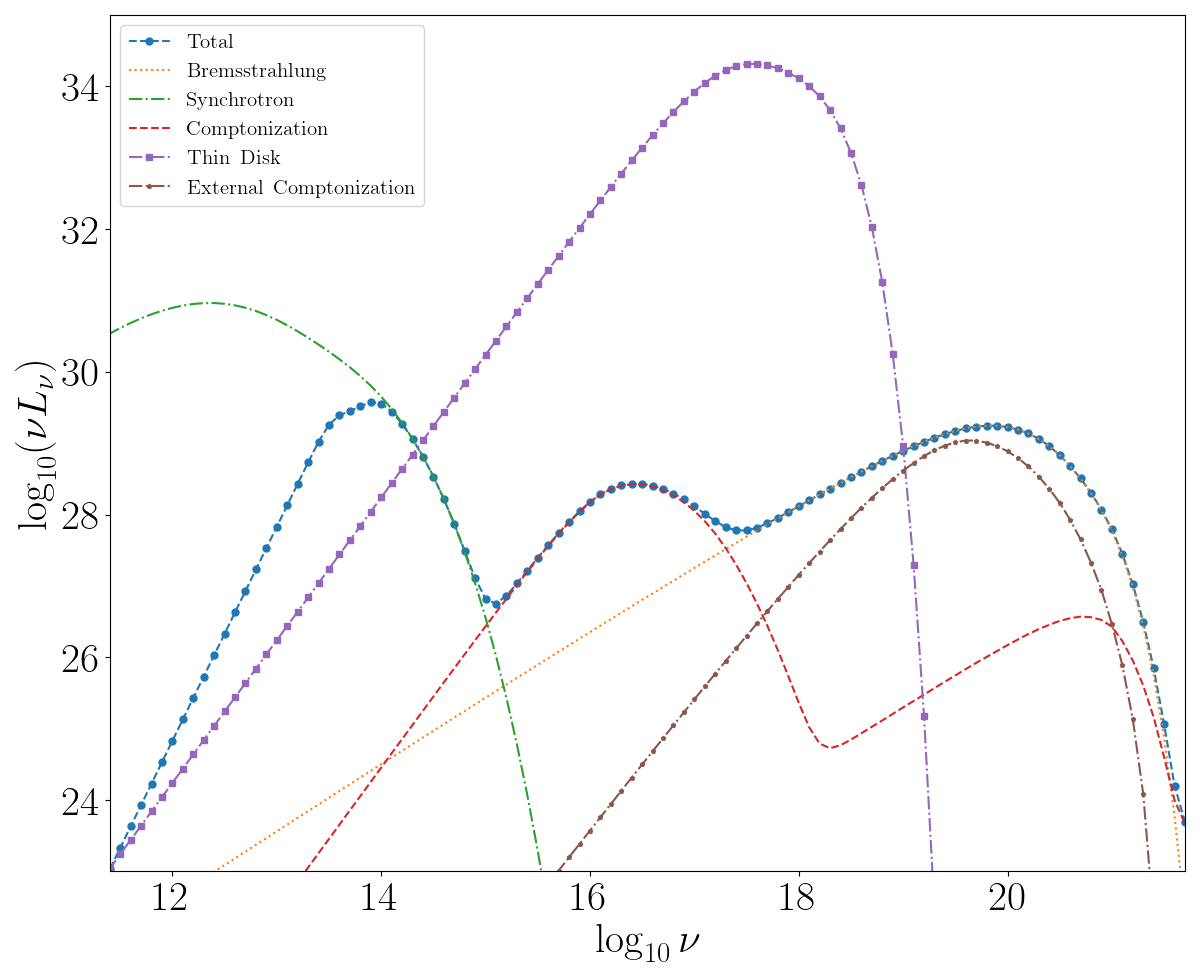}\label{sane}}
\subfloat[MAD]{
\includegraphics[width=0.49\textwidth]{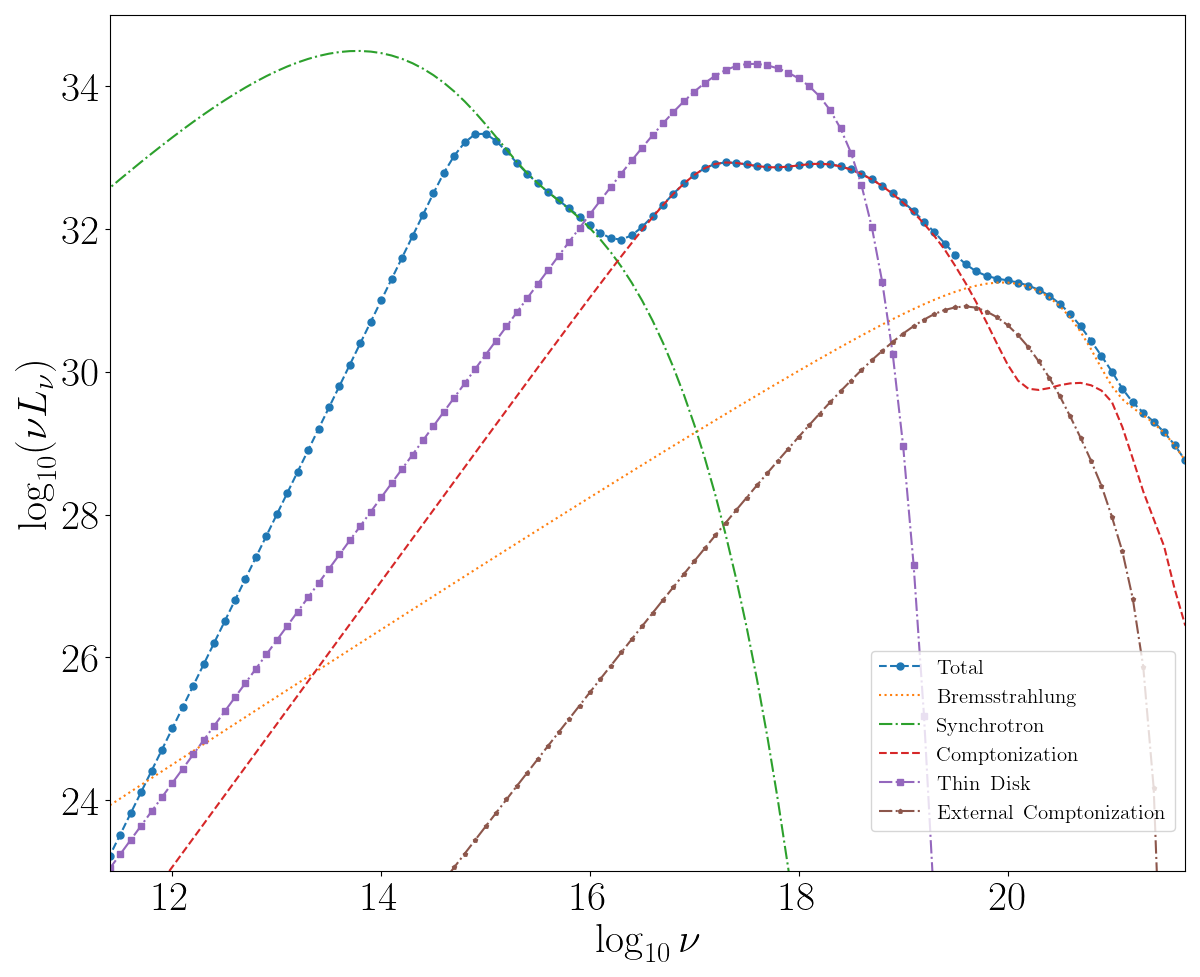}\label{mad}
}
     \caption{Components of spectra for MAD and SANE simulations for $a=0.9375$, $M=10M_\odot$ and $\dot{M}=10^{-3}\dot{M}_{Edd}$.}
     \label{sim-spec-comp}
 \end{figure*}

As evident from Table \ref{tab:sim-temp}, MADs show higher variations in $T_e$ as compared to SANEs. This is because MADs are hotter than SANEs and allow a higher range of physical $T_e$s in the accretion flow. We choose the $T_e$ values while maintaining a minimum $T_p/T_e$ ratio of 25. In SANEs, as the flows are cooler, the variation of $T_e$ at the outer edge $\gtrsim10^{-4}$ (in code units) leads to unphysical (or no) solutions.
 
The obtained $T_e$ and ion temperature ($T_p$) are shown in Figs.~\ref{temp-a5} and \ref{temp-a9}.  $T_p$ is higher than $T_e$ at least by 25 times (on average $T_p/T_e=81.73$ for SANE, and $729.48$ for MAD, for $a=0.5$; and $83.14$ for SANE, and $360.93$ for MAD, for $a=0.9375$) and the MAD flows are hotter than SANE flows for all the cases. $T_e$ for MAD and SANE flows are almost similar around the outer edge of the disk. MADs acquire a higher $T_e$ close to the black hole than SANEs. This is because the magnetic fields in MADs drop sharply near the black hole, leading to lower synchrotron cooling and thus an increase in $T_e$. For SANEs, the magnetic fields are almost constant throughout, leading to almost flat $T_e$ profiles as well.
\begin{figure*}
\centering
\subfloat[$a=0.5$]{
\includegraphics[width=0.49\textwidth]{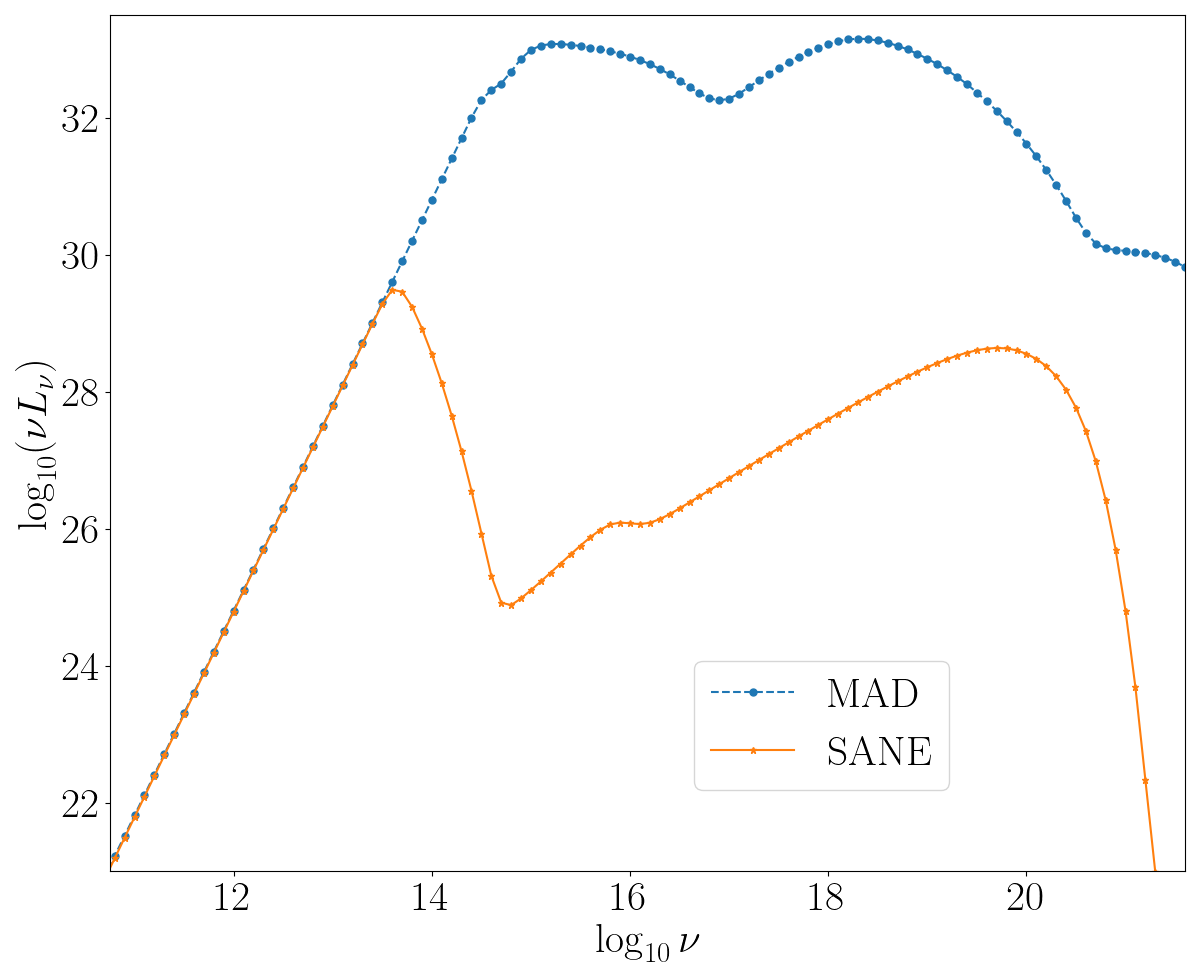}\label{sim-a5}}
\subfloat[$a=0.9375$]{
\includegraphics[width=0.49\textwidth]{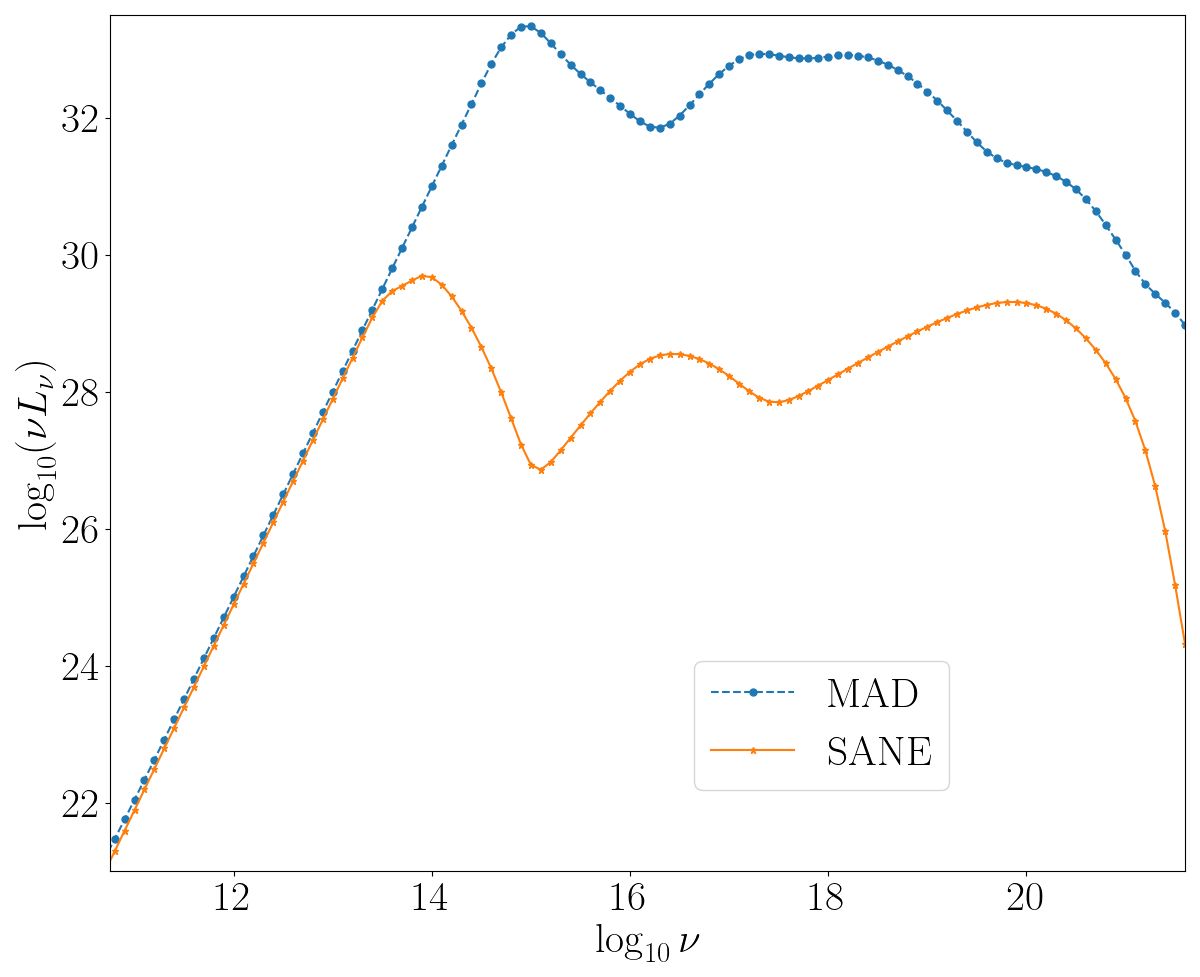}\label{sim-a9}
}
     \caption{Comparison of MAD and SANE spectra for $a=0.5$ and $a=0.9375$. Other disk parameters are same as in Fig.~\ref{sim-spec-comp}.}
     \label{sim-spec}
 \end{figure*}
\subsection{Spectra calculation}
We follow the approach described in $\S$~\ref{spec-cal} to evaluate the spectra for our simulations. The obtained spectra with their contributing components are shown in Fig~\ref{sim-spec-comp}. The spectra have contributions from synchrotron radiation, SSC and bremsstrahlung photons. For MAD simulations (Fig.~\ref{mad}), the synchrotron peak occurs around $\nu=10^{15}\text{Hz}$. The SSC has a double peaked feature. This is because of contributions to SSC luminosity from different parts of the disk due to the profiles like $T_e$, magnetic field etc. The bremsstrahlung peak follows the SSC peak. In our numerical steady state spectra (as in Fig.~\ref{spec1}), the EC peak overrides the bremsstrahlung peak. However, in our simulations, the density ($\rho$) is about an order of magnitude  higher than that of the numerical solutions (shown in Fig~\ref{rho}). As the bremsstrahlung flux goes as $\propto\rho^2$, the bremsstrahlung peak becomes larger than the EC peak in simulation based spectra.

Similar behavior is also obtained for SANE simulations shown in Fig~\ref{sane}. As the magnetic field is around an order of magnitude lower than in MAD, the synchrotron peak shifts to $\nu=10^{14}\text{Hz}$ and the peak reduces by three orders. The SSC peak also consequently shifts to $\sim10^{16}\text{Hz}$ and its luminosity also reduces by around four orders. The double peaked feature obtained for MAD is not present in this case, further underscoring the impact of flow profiles on the spectrum. The bremsstrahlung peak is larger than the EC peak, similar to the MAD case. However, the overall spectra get a small contribution from the EC peak at around $\nu=10^{20}\text{Hz}$.

We compare the MAD and SANE spectra for $a=0.5$ and $a=0.9375$ in Fig.~\ref{sim-spec}. The shapes of the spectra for both these cases are quite distinct. The bolometric luminosity in MAD simulations is higher than SANE by three orders of magnitude for both the spins. The bremsstrahlung peak is quite prominent in SANE spectra as the synchrotron peak is small due to low magnetic fields. This leads to lower SSC peaks, making the bremsstrahlung peak to appear quite prominently. 
The ratio of synchrotron to SSC luminosity is also less in the MAD case as compared to SANEs. This ratio turns out to be $\sim13.6$ and $\sim2.5$ for SANE and MAD for $a=0.9375$, respectively, and $\sim2.45\times10^3$ and $\sim0.83$ for SANE and MAD for $a=0.5$, respectively. For $a=0.5$ the SSC peak is slightly higher than the synchrotron peak in MAD, while for SANE, it barely contributes to the total spectra. The broadness of the SSC peak is also larger for MADs than SANEs. This is because $T_e$ is larger for MADs (albeit close to the black hole) and is similar to the SSC behavior as seen in Figs.~\ref{spec-alp} and \ref{spec-tec}.

Thus, the evolution of magnetic field and its structure in the disk strongly affects its spectra. Since the flow is sub-Keplerian and advective, the spectra just scales with the mass of the black hole and $\dot{M}$ of the flow. Hence, the diagnostics related to the peak positions and their relative heights in the spectra can indicate the magnetic field characteristics of the accretion flow.

\section{Conclusion} \label{conclusion}

In this work, we have systematically studied the various non-thermal emissions from a sub-Keplerian accretion flow around a rotating black hole and the variation of these emissions with various accretion disk properties. For SBHs, the peaks in the spectrum originate from synchrotron radiation (IR), SSC (X-ray) and EC ($\gamma$-ray) of soft photons from the Keplerian disk outside the sub-Keplerian region, with the synchrotron peak being the most prominent. For SMBHs, in addition to the emission mechanisms mentioned above, the bremsstrahlung peak also features at $\nu\sim10^{20}\text{Hz}$, with the highest being the EC peak. For SMBHs, the synchrotron and SSC peaks shift by three orders of magnitude in frequency and the EC shifts by one order. This showcases the fundamental differences between the spectra of SBHs and SMBHs.

Parameters like black hole spin $a$, mass accretion rate $\dot{M}$ and magnetic field strength $B$ also strongly affect the spectrum. The spectra peaks reduces and shift to the left with a decrease in $a$ as the energy and flux of the emitted photons decrease due to the interplay between electron temperature $T_e$, $B$ and density $\rho$. 
With the decrease of $\dot{M}$, the peaks of the spectra reduce, due to a reduction in the density of flow, leading to lower energy and lower flux photons. Reduction in $B$ leads to a reduction in the height of the synchrotron luminosity. As the disk size increases with lower $B$, the SSC and EC peaks increase due to more scattering by the higher amount of disk plasma. $T_e$ also has a strong impact on the spectral emission of the flow. Variations in $\alpha$, $T_e$ at the critical point $T_{ec}$, viscous and Ohmic heating fractions received by electrons, $x_f$ and $x_{fm}$, respectively, leading to an increase in $T_e$, lead to an increase in the bolometric luminosity and vice versa. The broadness of the SSC peak increases with the increase in $T_e$, indicating its usefulness in determining $T_e$. However, $x_f$ and $x_{fm}$ do not have much effect on the bolometric luminosity.
The magnetic field in our disks increases outwards, merging to the outer sub-Keplerian disk (not explicitly modeled here) enveloping central Keplerian disk. This field behavior leads to faster angular momentum transport due to magnetic shear. This, consequently, leads to a smaller sub-Keplerian disk, when the disk specific angular momentum $\lambda$ becomes $\lambda_K$ (Keplerian $\lambda$) at a shorter distance from the black hole (i.e. the outer Keplerian--sub-Keplerian ``sandwitch" region advances closer to the black hole). This affects the overall luminosity from the accretion flow. The presence of outflow, i.e. velocity along the $z$ direction ($v_z$), allows for a decreasing $B$ away from the black hole. This can lead to larger disk sizes and sustained high magnetic fields, particularly closer to the black hole, which will lead to higher luminosities \citep{ULX20}. 

To further explore the spectral variations with magnetic field structure in a more realistic system, we have also studied the spectral emissions from an accretion system around a rotating black hole simulated based on the GRMHD code \texttt{BHAC}. We have considered the SANE and MAD magnetic vector potentials to evolve a FM torus setup around the black hole. The resulting system is transformed to the locally non-rotating frame (LNRF), followed by time and density averaging of the resulting quantities. These transformed quantities are then used to calculate $T_e$ via the electron energy equation. MADs are found to have higher $T_e$ than SANE close to the black hole, due to a drop in magnetic field in this region, leading to lower synchrotron cooling. Farther away from the black hole ($r>5r_g$), $T_e$ becomes almost equal for MAD and SANE for both $a$ considered. As MADs have about an order of magnitude higher $B$ as compared to SANE, the bolometric luminosity of MADs is about three orders of magnitude higher. The synchrotron  and SSC peaks also shift by about two orders of magnitude between MAD and SANE. The bremsstrahlung peak dominates over the EC peak due to slightly higher density in our simulation systems. The SSC peak shows a double peaked structure in high spin MAD system, exhibiting the importance of emissions from different parts of the disk leading to distinct features in the spectrum. This property depends on the profiles of the various quantities used for spectra calculations. The ratio of synchrotron and SSC luminosities is much lower for MADs than SANEs for both $a=0.5$ and $a=0.9375$. This can be useful diagnostic for determining the magnetic field characteristics of the flow, i.e. whether the flow is MAD or SANE. This showcases the important result that even though the macroscopic properties of an accretion flow like mass of the black hole and accretion rate are same, the way the magnetic field evolves also leads to large changes in the spectrum of the sources, further underscoring the importance of magnetic field in the evolution and emission properties of accreting systems.

As flux eruption events are exaggerated in two-dimensions for MAD simulations, three-dimensional GRMHD simulations of MAD systems will offer a more accurate picture of the spectral features of the accretion flow. 
Note that our $T_e$ calculation is done in post-processing for both the 1.5-dimensional numerical steady state system and the GRMHD simulations. At low accretion rates, considered here, cooling becomes dynamically unimportant and, thus, the flow profiles would not be affected much by incorporating the electron energy equation in the flow solution. However, to study the system across different accretion rates, it would be beneficial to include the electron energy equation in the flow solutions and to solve it simultaneously for both the steady state and the GRMHD simulation calculations.

\section*{Acknowledgments}
The authors thank the referee for a thorough reading of the manuscript and many useful comments to improve the presentation.  
MP acknowledges the Prime Minister’s Research Fellows (PMRF) scheme for providing fellowship and IUCAA, Pune for hospitality where a part of this work was done. SG acknowledges IUCAA, Pune, for their support through the Visiting Associateship Programme. Additionally, BM would like to thank IUCAA, Pune for hospitality during a visit where a part of the work was done and their Visiting Associateship Programme.

\bibliography{Reference}{}
\bibliographystyle{aasjournalv7}

\end{document}